\documentclass[lettersize,journal]{IEEEtran}
\usepackage{amsmath,amsfonts}
\usepackage{algorithmic}
\usepackage{algorithm}
\usepackage{array}
\usepackage[caption=false]{subfig}
\usepackage{textcomp}
\usepackage{stfloats}
\usepackage{url}
\usepackage{verbatim}
\usepackage{fvextra}
\usepackage{graphicx}
\usepackage{cite}
\usepackage{comment}
\usepackage{tcolorbox}
\usepackage[hidelinks,hypertexnames=false]{hyperref}
\usepackage[nameinlink,capitalize,noabbrev]{cleveref}
\usepackage{autonum}
\usepackage[dvipsnames]{xcolor}
\newcommand{\okada}{\textcolor{black}}

\tcbset{
  sharpbox/.style={
    colback=white,          
    colbacktitle=white,     
    colframe=black,         
    coltitle=black,         
    arc=0pt,                
    outer arc=0pt,          
    boxrule=0.8pt,          
    fonttitle=\bfseries,
  }
}

\begin{document}

\title{Transforming Keystroke Noise to Text: \\Self-Supervised Acoustic Eavesdropping Attacks on Keyboards}

\author{Atsunori Okada, 
        Akira Ito, 
        Rei Ueno, ~\IEEEmembership{Member,~IEEE,}\\
        Yuichi Hayashi, ~\IEEEmembership{Senior Member,~IEEE,} 
        and~Naofumi Homma, ~\IEEEmembership{Senior Member,~IEEE}
\thanks{This work was supported in part by JST K Program Grant No.~JPMJKP24C1, Japan.}
\thanks{Atsunori Okada is with the Graduate School of Engineering, Tohoku University, 980-8579, Miyagi,  Japan (e-mail: okada.atsunori.p8@dc.tohoku.ac.jp).}
\thanks{Akira Ito and Naofumi Homma are with the Research Institute of Electrical Communication, Tohoku University, 980-8577, Miyagi, Japan (e-mail: akira.ito.b1@tohoku.ac.jp; naofumi.homma.c8@tohoku.ac.jp).}
\thanks{Rei Ueno is with the Graduate School of Informatics, Kyoto University, 606-8501, Kyoto, Japan (e-mail: ueno.rei.2e@kyoto-u.ac.jp).}
\thanks{Yuichi Hayashi is with the Graduate School of Science and Technology, Nara Institute of Science and Technology, 630-0912, Nara, Japan (e-mail: yu-ichi@is.naist.jp).}
}


\IEEEpubid{}

\maketitle

\begin{abstract}
We present a self-supervised acoustic eavesdropping attack that reconstructs typed text solely from keystroke sounds, without requiring labeled data for the target device.
The proposed attack enables stealthy eavesdropping in two real-world settings: (semi-)public physical spaces and online meetings.
Our method combines unsupervised acoustic clustering with Transformer-based language model inference and iterative self-training, enabling stable character inference under highly uncertain acoustic-to-character mappings. 
We demonstrate that, across four laptops and three input texts, the proposed method achieves a mean Levenshtein score of 0.941 with only 150 observed keystrokes in the nearby smartphone-recording setting, substantially outperforming prior unsupervised baselines in low-data regimes.
We further evaluate the method under several challenging acquisition conditions, including recording from approximately 3.0\,m away on the same desk, through-the-wall eavesdropping with a contact microphone, and keystroke sounds transmitted as background audio during online meetings.
Across these scenarios, the proposed method often achieves accuracy exceeding 0.9 with approximately 150--250 observed keystrokes.
These results demonstrate that accurate text reconstruction remains possible under the evaluated audio-only conditions, even with limited observed keystrokes and without requiring device-specific labeled data, highlighting a realistic and previously underestimated privacy risk.

\end{abstract}

\begin{IEEEkeywords}
Keyboard eavesdropping, Acoustic side-channel, Self-supervised learning.
\end{IEEEkeywords}

\section{Introduction}

\subsection{Background}
\label{sec:background}

\IEEEPARstart{R}{ecovering} typed text \okada{from physical leakage} poses a serious threat to user privacy and security because it can expose private communications, business discussions, and confidential documents.
People routinely type on laptops in public and semi-public spaces, including cafes, libraries, public transportation, and waiting lounges.
Participants in online meetings may also type while active microphones capture keystroke sounds.
These settings create opportunities for an attacker to exploit physical leakage from typing without compromising the target device. 
In this work, we ask how far keystroke eavesdropping can progress toward opportunistic text reconstruction under realistic attacker assumptions. 
We consider such an attack practical and powerful only if it satisfies three properties: \emph{(i) minimal deployment constraints}, \emph{(ii) no labeled data from the target device}, and \emph{(iii) high reconstruction accuracy from only a limited number of observed keystrokes}.

\okada{Prior work has demonstrated keystroke leakage across multiple physical side channels, but no existing attack jointly satisfies these three requirements.}
Acoustic attacks exploit key-dependent differences in keystroke sounds, but typically require target-specific labeled recordings or, in unlabeled settings, sufficiently many observations and favorable linguistic structure for reliable reconstruction~\cite{asonov2004keyboard,zhuang05:keyboard-acoustic,harrison2023practical,ayati25:llm-typo-correction,berger2006dictionary,furst2025practical}. 
Electromagnetic and wireless attacks recover keystrokes from keyboard emanations or wireless channel variations, but depend on dedicated sensing equipment, compatible infrastructure, and favorable sensor placement~\cite{vuagnoux2009compromising,ali2015keystroke,li2016csi,yang22:wireless-training-free,hu2023password}.
Vision-based attacks infer typed input from hand and finger motions, but require visual access to the target and sufficiently favorable camera placement~\cite{yue2015blind,shukla2014beware,chen2018eyetell}. 
Thus, although these modalities demonstrate that typing leaks substantial physical information, existing attacks remain constrained by profiling requirements, specialized sensing or infrastructure, favorable observation conditions, or the need for large amounts of input.

\IEEEpubidadjcol

Among these modalities, acoustic leakage is particularly well suited to opportunistic attacks because it imposes minimal deployment constraints, despite the limited reconstruction accuracy achieved by existing unlabeled attacks.
Keystroke sounds can be captured passively with commodity or portable recording devices, without device modification, specialized infrastructure, or line of sight, and may also leak through microphones used for online communication.
If such observations also permit accurate reconstruction without target-specific labeled data and from only a limited number of keystrokes, acoustic eavesdropping would constitute a particularly practical attack vector.

The limited accuracy of existing unsupervised acoustic attacks may reflect decoder limitations rather than insufficient key-specific information in keystroke sounds.
Supervised acoustic attacks can learn the mapping from keystroke sounds to key identities directly from labeled recordings and achieve high recognition accuracy, but their dependence on target-specific or closely matched training data fails to satisfy the second requirement defined above~\cite{asonov2004keyboard,harrison2023practical,ayati25:llm-typo-correction}.
Without such labels, an attacker must infer recurring acoustic classes corresponding to individual keys and determine which character each class represents from the observed sequence.
Existing unsupervised attacks typically address these problems by clustering acoustically similar keystrokes and resolving the resulting cluster-to-character correspondence using hidden Markov models (HMMs), dictionaries, or related statistical and lexical constraints~\cite{zhuang05:keyboard-acoustic,berger2006dictionary,furst2025practical}.
These decoders provide substantially weaker contextual constraints than modern neural language models, which may limit their ability to resolve cluster-to-character ambiguity under sparse observations.
Consequently, the limited reconstruction performance reported by existing unsupervised acoustic attacks may reflect the capability of their inference methods rather than a fundamental limit of the acoustic leakage itself.

Modern Transformer-based language models provide stronger contextual inference and may therefore overcome this decoding bottleneck.
Recent acoustic work has used a large language model (LLM) to correct character strings after acoustic decoding~\cite{ayati25:llm-typo-correction}, but such post-hoc correction receives only the decoder's hard output and cannot reconsider uncertainty in the underlying correspondence between acoustic clusters and characters.
A more fundamental approach is to incorporate modern contextual inference directly into the acoustic-to-text decoding process, allowing linguistic context to help resolve uncertain cluster-to-character mappings themselves. 
If this enables reliable reconstruction from substantially fewer unlabeled observations, prior evaluations based on weaker decoders may have underestimated the practical risk of acoustic keystroke leakage. 
We therefore investigate \emph{how far acoustic eavesdropping can progress toward opportunistic text reconstruction when modern language-model inference is integrated directly into the decoding process}.

\subsection{Our Contribution}
\label{sec:contribution}
Our key idea is to preserve uncertainty in the cluster--character mapping and resolve it with contextual language inference before committing to hard character decisions.
Starting from unlabeled keystroke audio, we cluster acoustically similar keystrokes while maintaining, for each cluster, a probability distribution over candidate characters.
We convert this distribution into an ambiguity-aware character embedding and feed the resulting sequence to a character-level BERT, allowing bidirectional linguistic context to resolve cluster--character ambiguity before a hard character decision is made.
An LLM subsequently corrects residual contextual errors, and high-confidence predictions are propagated through the acoustic feature space and fed back to update the cluster--character mapping.
Repeating this process allows acoustic similarity and linguistic context to reinforce each other, enabling text reconstruction from limited observations without labeled acoustic data from the target device.

Our experiments test whether modern contextual decoding can extract substantially more useful information than conventional unsupervised decoders from the same type of acoustic representation.
Under controlled nearby smartphone recording, we compare the proposed method with HMM- and dictionary-based decoders, both followed by LLM correction, across four laptop platforms and three input texts.
We quantify reconstruction accuracy using a normalized Levenshtein score ranging from 0 to 1, where 1 denotes an exact match between the reconstructed and original text.
With 150 observed keystrokes, the proposed method achieves a mean Levenshtein score of 0.941 over the 12 device--text combinations, compared with 0.397 and 0.311 for the two baselines, respectively.
A separate ablation study shows that BERT with feedback can reconstruct text without an LLM, reaching 0.8831 at 400 observations, while LLM guidance improves accuracy and stability with fewer observations.
Across more challenging acquisition channels, the proposed method also reconstructs text accurately in conditions where the HMM- and dictionary-based methods remain unsuccessful.
Additional experiments evaluate robustness to controlled Gaussian and speech-babble noise and show that Shift-based capitalization and Backspace-based editing can be reconstructed within the same self-supervised framework.
Together, these results show that conventional or post-hoc decoding can substantially underestimate the attack capability available from keystroke acoustics under limited observations.

Our main contributions are:
\begin{itemize}
    \item We introduce, to the best of our knowledge, the first keyboard acoustic side-channel attack that integrates a Transformer-based language model directly into acoustic-to-text decoding.
    \item We compare the proposed method with conventional decoders followed by LLM correction across four laptops and three texts, and separately evaluate the contributions of iterative feedback and LLM guidance, demonstrating improved sample efficiency and reconstruction stability.
    \item We demonstrate the framework's robustness and extensibility through controlled-noise evaluation and an operation-aware extension that jointly infers Shift-press, Shift-release, and Backspace events.
\end{itemize}

\subsection{Scope and Other Modalities}
\label{sec:scope}
\okada{We deliberately adopt an audio-only threat model because keystroke sounds can be acquired passively and opportunistically using commodity or portable recording devices, without compromising the target device or deploying dedicated infrastructure. 
Other physical side channels surveyed in \cref{sec:background} are complementary and could potentially be combined with the proposed framework, but none of our evaluations assumes access to another sensing modality.}

\section{Threat Model}
\label{sec:threat-model}

\subsection{Adversarial Goals}

The attacker's goal is to reconstruct English natural-language text typed by a victim using only recorded keystroke sounds. 
Sensitive keyboard input includes both non-linguistic strings, such as randomly generated passwords and credit card numbers, and natural-language content, such as private messages, business correspondence, meeting notes, and confidential documents. 
Although the proposed method does not target non-linguistic strings because it exploits linguistic structure, reconstructing sensitive natural-language content alone poses a meaningful privacy and security risk.

\subsection{Victim Conditions}
We assume a typical computer user typing English text.
We characterize the victim as follows:
\begin{itemize}
\item \textbf{Character set.}
The core evaluations in \cref{sec:eval,sec:real} consider text composed of 29 symbols: lowercase letters \texttt{a}--\texttt{z} (26 letters), space, comma (,), and period (.).
\Cref{sec:operation-aware} extends this set with Shift press, Shift release, and Backspace tokens.

\item \textbf{Input device.}
The victim uses a standard built-in laptop keyboard.
We evaluate multiple laptop platforms in \cref{sec:eval,sec:real}.

\item \textbf{Typing speed.}
We assume a  typing speed of up to 300 characters per minute (CPM). 
Empirical studies report that typical users type at approximately 30--60 words per minute~\cite{dhakal2018observations, pinet2022typing}, corresponding to about 150--300 CPM assuming five characters per word. 
This implies an average keystroke interval of roughly 200--400 ms. 
The proposed attack requires consecutive keystroke sounds to remain sufficiently separated for waveform segmentation.

\end{itemize}

\subsection{Adversarial Assumptions and Constraints}
The attacker is assumed to observe the audio waveform produced by the victim's typing but to have no other information.
The audio waveform is captured using an off-the-shelf microphone. 
To study a more deployment-oriented attack setting, we impose the following constraints:
\begin{itemize}
\item \textbf{Limited environmental control and portable recording setup.}
Depending on the situation, the attacker may be able to place a microphone, but cannot arbitrarily change the victim's keyboard or device configuration.
Although the microphone may be placed in a favorable position for low-noise recording, the attacker does not know its exact distance or angle from the target device.
\item \textbf{No prior knowledge, no profiling, and no compromise of the target device.}
The attacker has no detailed knowledge of the victim's keyboard model or of the type of English text being typed.
The attacker obtains neither the ground-truth typed text nor any labeled dataset that maps keys or characters to their corresponding acoustic signals.
The attacker does not use malware such as keyloggers, access OS-internal information, or abuse application privileges on the victim device.
\end{itemize}
Together, these constraints instantiate the three practical requirements introduced in \cref{sec:background}: an off-the-shelf and portable acquisition setup, no labeled acoustic data for the target, and text reconstruction from a limited number of observed keystrokes.

\subsection{Evaluated Acquisition Settings and Attack Scenarios}
We evaluate whether the attack remains viable across three practically distinct acoustic acquisition channels: airborne recording, structure-borne recording, and online-meeting audio.
We instantiate these channels using \textbf{(a) a portable voice recorder or smartphone}, \textbf{(b) a portable contact microphone}, and \textbf{(c) a laptop's built-in microphone}, with audio sampled at 44.1~kHz.

These setups represent three plausible eavesdropping scenarios considered in our evaluation: distant recording in \mbox{(semi-)public} spaces, through-the-wall recording using structure-borne vibrations, and online-meeting eavesdropping.
Setup (c) targets \textbf{online meetings}. 
In the online setting, the attacker captures keystroke sounds transmitted as background audio when noise suppression is disabled, unavailable, or ineffective.
Moreover, noise cancellation may be disabled by users or available only on certain paid plans.
Under such conditions, an attacker can exploit keystroke sounds leaked through meeting audio.

\section{Attack Design}\label{sec:proposed}

\subsection{Overview}

\begin{figure*}[t]
    \centering
    \includegraphics[width=0.90\linewidth]{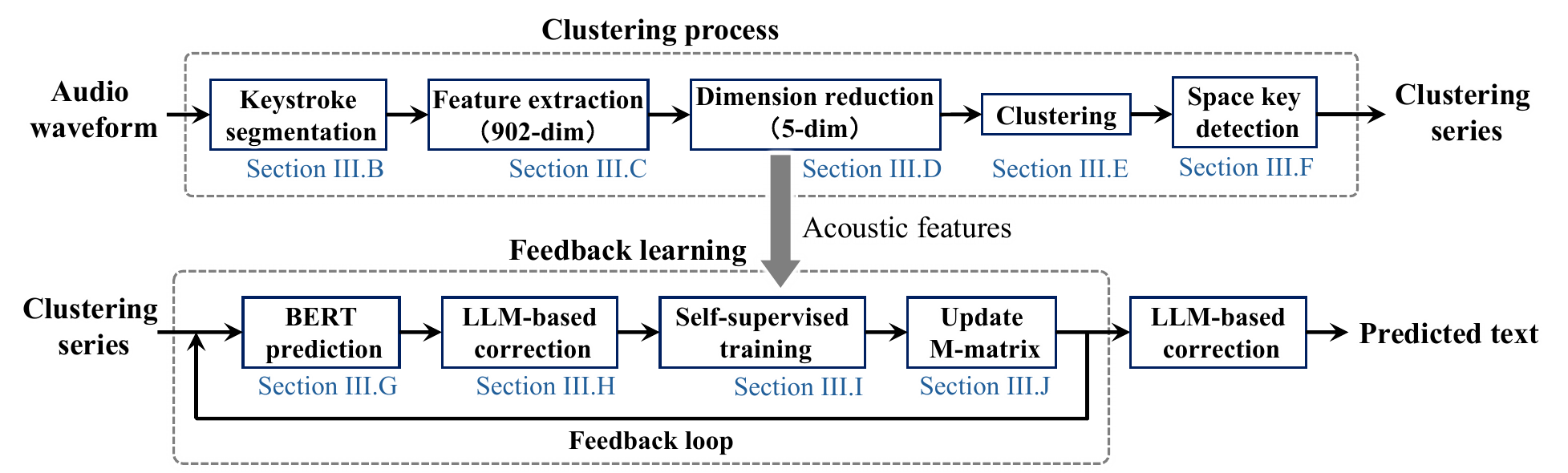}
    \caption{Proposed attack pipeline.}
    \label{fig:propose_block}
\end{figure*}

The proposed attack preserves uncertainty in the acoustic cluster--character mapping and iteratively resolves it by combining acoustic similarity with linguistic context. \cref{fig:propose_block} summarizes the resulting end-to-end pipeline.
Its acoustic front end detects and segments keystrokes (\cref{sec:segmentation}), extracts acoustic features (\cref{sec:feature}), and applies UMAP and hierarchical clustering (\cref{sec:umap,sec:clustering}). 
It also identifies acoustically distinctive space-key events, which provide useful anchors for contextual inference (\cref{sec:space}).
We represent each cluster by a probabilistic cluster--character mapping and transform that distribution into an ambiguity-aware embedding. 
A character-level BERT then resolves the mapping using bidirectional linguistic context before a hard character decision is made (\cref{sec:bert}). 
An LLM subsequently corrects residual contextual errors (\cref{sec:llm}); agreement between the two outputs supplies pseudo-labels, which label spreading propagates through the acoustic feature space (\cref{sec:labelspreading}). 
The resulting labels update the cluster--character mapping, and the decoder repeats this feedback loop to improve reconstruction in the low-observation regime (\cref{sec:feedback}).

\subsection{Keystroke Segmentation}
\label{sec:segmentation}

\begin{figure}[t]
  \centering
  \includegraphics[width=0.85\linewidth]{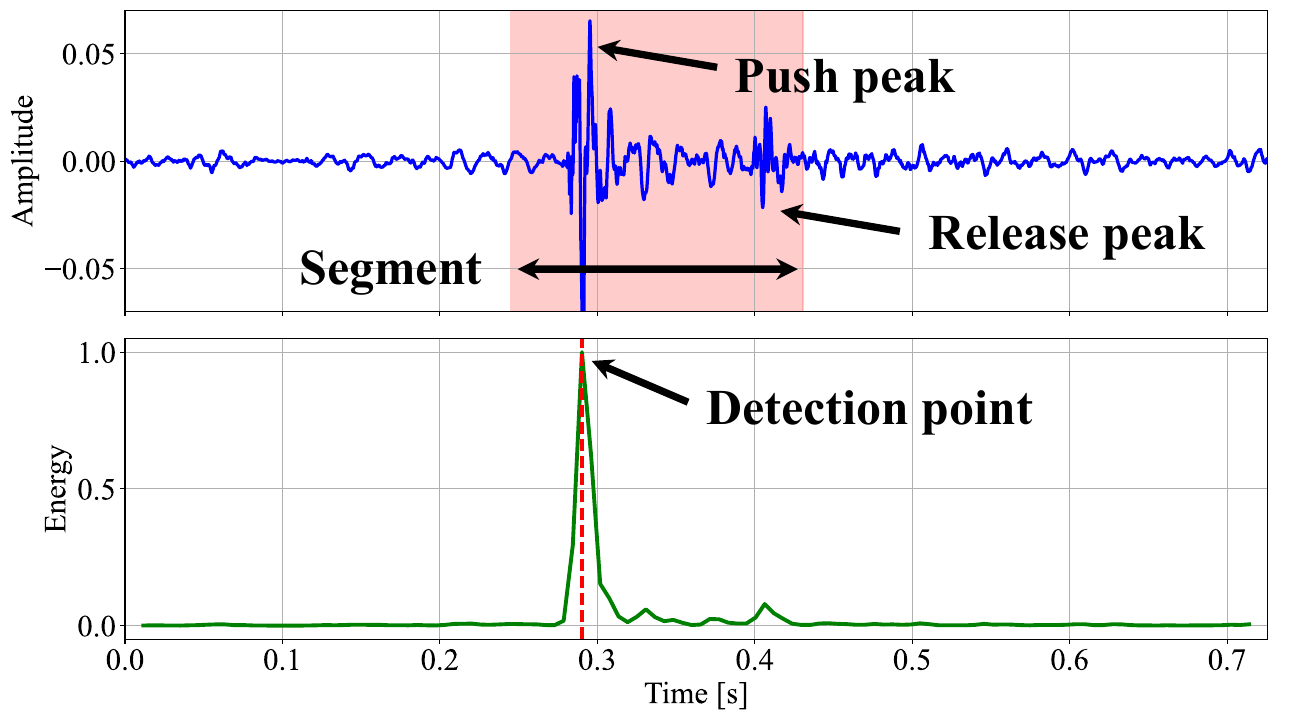}
  \caption{Example of audio segmentation. The top panel shows the waveform, and the bottom panel shows the short-window energy curve.}
  \label{fig:keydetection}
\end{figure}

To obtain one acoustic observation per keystroke for subsequent clustering, we detect individual events in the continuous waveform and extract fixed-length segments around them, as illustrated in \cref{fig:keydetection}.
Given an audio waveform $x(t)$, we compute an STFT with a window length of 1,024 samples and detect peaks in the resulting short-window energy using prominence-based peak detection~\cite{scipy-signal-find-peaks}.
Let $P = \{p_1, p_2, \dots, p_N\}$ be the detected peak indices, where $N$ is the number of detected keystrokes.
For each $p_i$, we extract an $L=8{,}192$-sample segment,
$\mathbf{s}_i = \left(x_{p_i-L_{\text{pre}}}, x_{p_i-L_{\text{pre}}+1}, \dots, x_{p_i+L_{\text{post}}-1}\right)$,
where $L_{\mathrm{pre}}=2{,}000$ and $L_{\mathrm{post}}=6{,}192$.

To verify and correct keystroke timing detection and segmentation,
we use a lightweight GUI to inspect the waveform and the detected peaks, and to add, remove, or adjust peak positions before re-extracting segments from the corrected peak set $P$.

\subsection{Feature Extraction}
\label{sec:feature}
\okada{We represent each segmented keystroke by a 902-dimensional acoustic feature vector that combines MFCC statistics, MFCC derivatives, and complementary spectral descriptors.}
First, for each segment $\mathbf{s}_i$, we compute mel-frequency cepstral coefficients (MFCCs), a feature representation widely used in speech recognition and audio analysis.
We set the FFT window length to 4,096 samples, the hop length to 512 samples, and the number of mel bands to 128.
The MFCC features extracted from a segmented waveform $\mathbf{s}_i\in \mathbb{R}^{8{,}192}$ are represented by $\operatorname{MFCC}(\mathbf{s}_i) \in \mathbb{R}^{128 \times F}$, where $F$ denotes the number of time frames.
Let $\mathbf{m}_{i,j} \in \mathbb{R}^{F}$ be the $j$-th coefficient trajectory.
We compute summary statistics $\mathbf{m}_i^{\mathrm{mean}},\ \mathbf{m}_i^{\mathrm{var}},\ \mathbf{m}_i^{\mathrm{max}},\ \mathbf{m}_i^{\mathrm{min}}$, and $\ \mathbf{m}_i^{\mathrm{med}} \in \mathbb{R}^{128}$, which correspond to the mean, variance, maximum, minimum, and median over time for $\operatorname{MFCC}(\mathbf{s}_i)$, respectively.
In addition, we define a spectral feature vector $\mathbf{z}_i \in \mathbb{R}^{6}$ that concatenates the mean and variance of the spectral centroid, spectral rolloff, and zero-crossing rate.
We also compute MFCC derivative features $\Delta_i,\, \Delta^2_i \in \mathbb{R}^{128}$.
The resulting 902-dimensional feature vector is
\begin{equation}
\mathbf{f}_i = \bigl[\ \mathbf{m}_i^{\mathrm{mean}}, \mathbf{m}_i^{\mathrm{var}}, \mathbf{m}_i^{\mathrm{max}}, \mathbf{m}_i^{\mathrm{min}}, \mathbf{m}_i^{\mathrm{med}}, \mathbf{z}_i, \Delta_i, \Delta^2_i \bigr]^\top.
\end{equation}
We standardize these feature vectors before dimensionality reduction.

\subsection{Dimensionality Reduction}
\label{sec:umap}
In this work, we target eavesdropping attacks that operate with only a limited number of observations (e.g., 100--200 keystrokes), where reliable inference must be achieved under severe data constraints.
In this regime, the feature dimensionality (902) is substantially larger than the number of observations.
This imbalance may make the feature space sparse and degrade clustering performance.
Therefore, we apply UMAP~\cite{mcinnes2018umap, UMAPUnif63:online} to map each feature vector from 902 dimensions to 5 dimensions.
Let $\phi_{\mathrm{UMAP}} : \mathbb{R}^{902} \rightarrow \mathbb{R}^{5}$ be the UMAP mapping.
Then, the low-dimensional representation for each keystroke sample is
$\mathbf{u}_i = \phi_{\mathrm{UMAP}}(\mathbf{f}_i) \in \mathbb{R}^{5},~ i = 1,2,\dots,N$,
which is used as the input to the hierarchical clustering step in \cref{sec:clustering}.

\subsection{Clustering}
\label{sec:clustering}
We cluster the low-dimensional representations to recover recurring acoustic classes without requiring key labels.
We apply agglomerative clustering~\cite{ward1963hierarchical,lance1967general,kaufman2009finding} to the UMAP representations $\mathbf{u}_i$ to group acoustically similar keystrokes.

Because UMAP places acoustically similar samples close to each other in the embedded space, hierarchical clustering enables us to assign acoustically similar keystroke sounds to the same cluster. 
\okada{Rather than fixing the number of clusters, each initialization run in \cref{sec:bert} draws
$K$ randomly from $20\pm5$ (i.e., 15--25), allowing different cluster granularities under different observation counts.}
After clustering, the cluster index assigned to the $i$-th keystroke sound sample is denoted by $c_i \in \{1,2,\ldots,K\}$, and the resulting cluster-index sequence is
$\mathbf{c} = (c_1, c_2, \ldots, c_N)$.

\subsection{Space-Key Identification}
\label{sec:space}

The space key often produces an acoustic signature distinct from those of letter keys because of its distinctive size, shape, and typing motion~\cite{zhuang05:keyboard-acoustic}.
We exploit this property to identify space keystrokes in advance and use their positions as reliable anchors for the subsequent language-model-based decoding, which substantially improves stability.

\paragraph{Seed selection}
We first manually select a single space-key sample by listening to the segmented keystroke sounds.  
In practice, the space keystroke is easy to distinguish by ear in our recordings, and selecting one representative sample requires only minimal manual effort.

\paragraph{Distance-based identification in the UMAP space}
Let $\mathbf{u}_i \in \mathbb{R}^{5}$ denote the UMAP embedding of the $i$-th keystroke (\cref{sec:umap}).
Given a manually selected seed space-key sample $s_{i^\star}$ with embedding $\mathbf{u}_{i^\star}$, we identify all keystrokes whose Euclidean distance to the seed is within a fixed threshold $\tau_{\text{space}} = 2.0$ as space keystrokes:
\begin{align}
\mathcal{S} = \left\{\, i \ \middle|\ \lVert {\mathbf u}_i - {\mathbf u}_{i^\star} \rVert_2 \le \tau_{\text{space}}\,\right\}.
\end{align}
We regard $\mathcal{S}$ as the set of space positions in the keystroke sequence.

\begin{figure}[t]
\centering
\subfloat[Ground-truth key labels]{\includegraphics[width=0.48\linewidth]{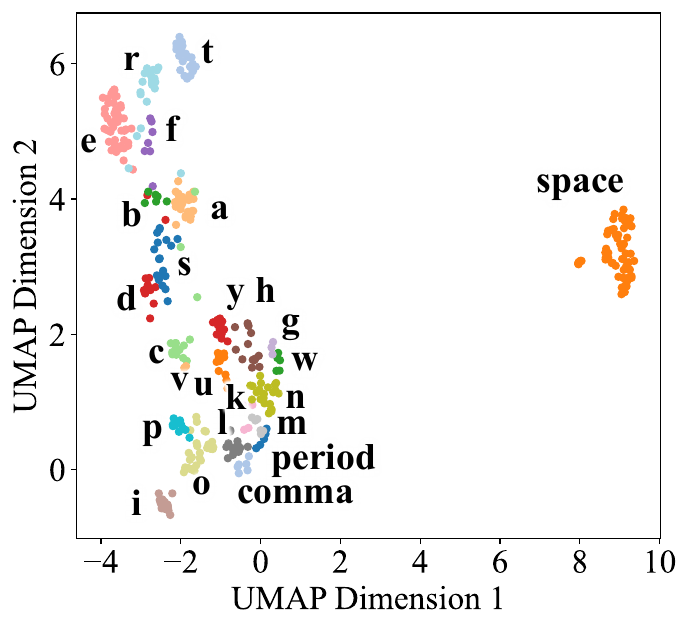}%
\label{fig:spacekey_identification-a}}
\hfil
\subfloat[Space-key identification]{\includegraphics[width=0.48\linewidth]{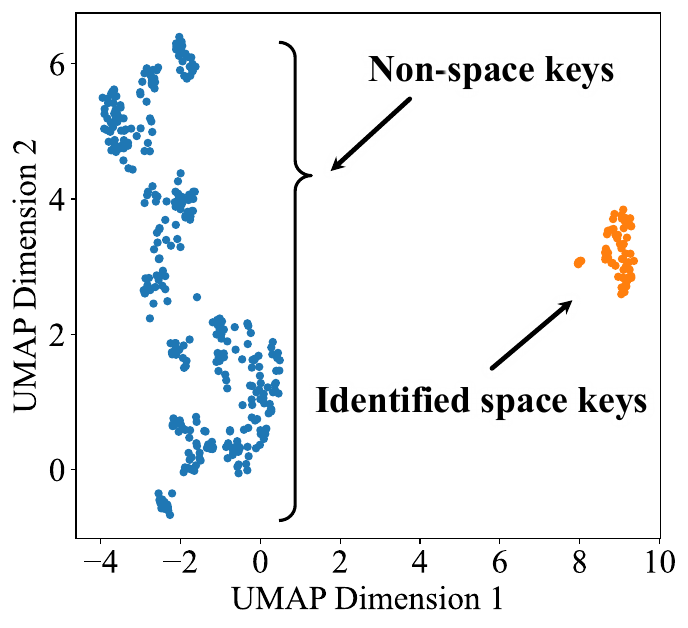}%
\label{fig:spacekey_identification-b}}
\caption{2-D visualization of keystroke distribution after UMAP-based dimensionality reduction.}
\label{fig:spacekey_identification}
\end{figure}

\paragraph{Visualization}
\cref{fig:spacekey_identification} visualizes an example of the keystroke-sound distribution in a 2-D UMAP embedding, where \cref{fig:spacekey_identification-a} shows the ground-truth key labels and \cref{fig:spacekey_identification-b} shows the result of the proposed method.
In \cref{fig:spacekey_identification-a}, space keystroke sounds form a compact cluster that is well separated from non-space keystroke sounds.
As a result, selecting a single seed sample and applying the distance threshold enables accurate identification of the remaining space keystrokes, as shown in \cref{fig:spacekey_identification-b}.

\subsection{BERT-Based Text Prediction}
\label{sec:bert}
We use a character-level BERT model~\cite{devlin2019bert} with masked language modeling (MLM) to resolve uncertainty in the cluster--character correspondence using bidirectional context.

\paragraph{Training a character-level BERT}
We train a custom character-level BERT on an OpenWebText-derived corpus~\cite{Gokaslan2019OpenWeb}.
The core model uses the 29-symbol vocabulary defined in \cref{sec:threat-model}: lowercase letters, space, comma, and period.
\Cref{sec:operation-aware} separately evaluates an extension with Shift and Backspace operations.
Training and model details are provided in \cref{ap:bert}.

\paragraph{Probabilistic cluster--character mapping}
We define a probabilistic correspondence between clusters and characters by a matrix
$\mathbf{M} \in [0,1]^{|V|\times K}$,
where $K$ is the number of clusters and $V$ is the vocabulary.
The entry $M_{j, c_i}$ represents the probability that the cluster $c_i$ (assigned to the $i$-th keystroke) corresponds to the $j$-th character $v^{(j)} \in V$.
We obtain an initial estimate $\mathbf{M}^{(0)}$ using an HMM-based procedure with the expectation--maximization (EM) algorithm~\cite{zhuang05:keyboard-acoustic}.

\paragraph{Constructing ambiguity-aware embeddings}
Transformer-based models convert each input token into an embedding vector and process the embedding sequence via attention.
Let $\mathbf{e}(v^{(j)}) \in \mathbb{R}^{d_e}$ denote the embedding vector for character $v^{(j)}$, where $d_e$ is the embedding dimension.
Because a cluster may correspond to multiple characters probabilistically through $\mathbf{M}$, we represent this uncertainty as an \emph{ambiguity-aware} embedding.
Specifically, we define the ambiguity-aware embedding for the $i$-th keystroke as
\begin{equation}
    \label{eq:aimai}
    \mathbf{h}_i = \sum_{j=1}^{|V|} \mathbf{M}_{j,c_i}\,\mathbf{e}(v^{(j)}).
\end{equation}
This weighted sum captures the uncertainty over which character corresponds to a given cluster.
By applying \cref{eq:aimai} to all positions, we obtain an embedding sequence
$\mathbf{H} = (\mathbf{h}_1, \mathbf{h}_2, \ldots, \mathbf{h}_N)$.

\paragraph{Character prediction with MLM}
We apply MLM to the embedding sequence $\mathbf{H}$ to predict characters.
For each position $i$, we create an input sequence where the ambiguity-aware embedding at position $i$ is replaced by \texttt{[MASK]}:
$\mathbf{H}^{(i)} = (\mathbf{h}_1,\mathbf{h}_2,\ldots,\mathbf{h}_{i-1},\mathbf{e}_{\texttt{[MASK]}},\mathbf{h}_{i+1},\ldots,\mathbf{h}_N)$.
The predicted character is then given by
\begin{equation}
\hat{y}_i
=
\arg\max_{v^{(j)} \in V}
P_{\mathrm{BERT}}\!\left(v^{(j)} \mid \mathbf{H}^{(i)}\right).
\end{equation}
We repeat this procedure for all positions $i=1,2,\ldots,N$ to obtain the final predicted character sequence
$\hat{\mathbf{y}} = (\hat{y}_1, \hat{y}_2, \ldots, \hat{y}_N)$.
Because BERT can exploit bidirectional context, it can resolve ambiguities where a single cluster may correspond to multiple characters and thereby improve prediction accuracy.
\okada{For sequences longer than the model's maximum input length, we apply BERT inference using overlapping sliding windows.}

\paragraph{Multiple initialization runs for BERT decoding}
Because UMAP and clustering are initialization-sensitive, we perform $R=10$ independent runs, with $K$ sampled as described in \cref{sec:clustering}.
Let $\hat{\mathbf{y}}^{(r)}=(\hat{y}^{(r)}_1,\ldots,\hat{y}^{(r)}_N)$ denote the decoded character sequence obtained from the $r$-th UMAP-and-clustering initialization run, and let
$P^{(r)}_{\mathrm{BERT},i}(v)$ be the BERT posterior probability of character $v$ at position $i$ in that run.
Since ground-truth text is unavailable, we score each initialization using the average log-likelihood of its own prediction:
\begin{equation}
    \label{eq:bert-init-score}
    \mathcal{L}^{(r)}
    =
    \frac{1}{N}
    \sum_{i=1}^{N}
    \log
    P^{(r)}_{\mathrm{BERT},i}\!\left(\hat{y}^{(r)}_i\right).
\end{equation}
We then select the initialization run
\begin{equation}
    \label{eq:bert-init-select}
    r^\star = \mathop{\mathrm{arg\,max}}\limits_{r \in \{1,2,\dots,R\}} \mathcal{L}^{(r)},
\end{equation}
and use the corresponding clustering result and decoded sequence as the initial state for the subsequent feedback loop.
This procedure chooses the most likely initialization under the BERT model itself, without requiring any labeled text from the target.

\subsection{LLM-Based Correction}
\label{sec:llm}
The BERT prediction $\hat{\mathbf{y}}$ can retain errors caused by an imperfect cluster--character mapping or ambiguous context. 
We therefore use an LLM as a complementary post-editing stage; unlike BERT, it does not resolve acoustic ambiguity directly but instead improves global linguistic consistency and provides evidence for subsequent self-training.

Let  $\mathrm{LLM}(\cdot)$ denote the LLM-based correction function.
Given a BERT prediction $\hat{\mathbf{y}}$, the corrected string is obtained as
$\mathbf{y}^\ast = \mathrm{LLM}(\hat{\mathbf{y}})$.
While BERT predicts characters from bidirectional context within its input window, an LLM can model longer-range dependencies across the entire passage and produce a grammatically and semantically consistent string.
In our implementation, we use Google Gemini 3.1 Flash Lite (version: \texttt{gemini-3.1-flash-lite}) to correct contextual errors in the BERT output and improve reconstruction accuracy.

\subsection{Self-Training with Label Spreading}
\label{sec:labelspreading}
We convert agreement between acoustic-contextual decoding and LLM correction into pseudo-labels and propagate these labels through the acoustic feature space for self-supervised refinement.

\paragraph{Confidence-based self-labeling}
We extract reliable pseudo-labels from spans on which the BERT prediction $\hat{\mathbf{y}}$ and the LLM-corrected sequence $\mathbf{y}^\ast$ sufficiently agree.
We align the two sequences at the phrase level using a whitespace-based diff.
For each aligned phrase pair $(o_m,c_m)$, where $o_m$ and $c_m$ denote the phrases before and after LLM-based correction, respectively, we compute a normalized edit similarity $\mathrm{sim}(o_m,c_m)\in[0,1]$ and regard the pair as high-confidence when it satisfies $\mathrm{sim}(o_m,c_m)\ge\tau_{\text{sim}}$ and $|o_m|=|c_m|$, where $\tau_{\mathrm{sim}}=0.6$.
For these phrases, we replace $o_m$ with $c_m$ to obtain a pseudo-labeled string $\tilde{\mathbf{y}}$. 
We also assign a confidence mask $m_i\in\{0,1\}$ to each character position $i$, where $m_i=1$ indicates that the position is covered by a high-confidence phrase and is treated as pseudo-labeled.

\paragraph{Semi-supervised refinement via label spreading}
We propagate the pseudo-labels to the remaining keystrokes by \emph{label spreading}.
For each keystroke index $i$, let $\mathbf{u}_i\in\mathbb{R}^5$ denote the feature vector obtained by UMAP, and let $\tilde{y}_i$ be its pseudo-character label when available.
Based on the confidence mask $m_i$, we partition the samples into a labeled set ($m_i=1$) and an unlabeled set ($m_i=0$).
Starting from the labeled set, we apply label spreading to infer character labels for all samples. 
Unlike hard self-labeling, label spreading allows labels to change during propagation when the local neighborhood structure supports a different label.
This flexibility helps the refinement process avoid poor local optima caused by early mistakes in $\tilde{\mathbf{y}}$.
By applying label spreading to all positions, we obtain a refined character sequence $\bar{\mathbf{y}}=(\bar{y}_1,\bar{y}_2,\dots,\bar{y}_N)$.

\subsection{Iterative Feedback}
\label{sec:feedback}
The feedback loop uses refined textual predictions to update the cluster--character mapping, allowing acoustic similarity and linguistic context to reinforce each other iteratively.
Using the refined character sequence obtained after self-training and the cluster sequence $\mathbf{c}$, we re-estimate a cluster--character correspondence matrix $\mathbf{M}'$.
We then update the current matrix $\mathbf{M}$ in a gradual manner using a mixing parameter $\beta$.
The update rule is given by
\begin{equation}
    \label{eq:updateM}
    \mathbf{M}_{\text{new}} = (1-\beta)\, \mathbf{M} + \beta\, \mathbf{M}'.
\end{equation}
Here, $\beta \in (0,1)$ is an update weight analogous to a learning rate, and we set $\beta = 0.8$ in our experiments.

Using the updated $\mathbf{M}_{\text{new}}$, we rerun BERT inference, LLM-based correction, and self-training, and repeat this feedback loop multiple times.
In our implementation, we use \(T_{\max} = 50\) iterations. 
We empirically found that 50 iterations were sufficient for stable convergence; therefore, we use this fixed value consistently across all experiments.

\section{Evaluation under Controlled Conditions}\label{sec:eval}
\okada{This section evaluates how effectively different decoders reconstruct text from the same acoustic observations under controlled nearby-recording conditions.
We compare the proposed method with existing unsupervised decoders across four laptop platforms and three input texts, and separately evaluate the contributions of iterative feedback and LLM guidance to reconstruction accuracy and sample efficiency.}

\subsection{Experimental Setup}
\label{sec:eval-setup}
\okada{We use a controlled nearby-recording setup and normalized edit-distance accuracy to isolate decoder performance under favorable acoustic conditions.}

\paragraph{Recording setup}
We use four off-the-shelf laptops: a Dell Latitude 7320, a 13-inch MacBook Pro (2019), an HP OmniBook X 14, and a Lenovo ThinkPad X390.
For each laptop, the victim types three English passages of 426, 426, and 438 characters, respectively, while a smartphone (Apple iPhone 15) placed beside the target laptop records the keystroke sounds.
The passages are business-like email messages written with the character set defined in \cref{sec:threat-model} and contain simple proper nouns such as personal names.
The resulting 12 recordings provide a controlled cross-platform baseline for isolating decoder performance before we turn to the more challenging acquisition conditions in \cref{sec:real}.

\paragraph{Evaluation metrics}
\label{sec:eval-metrics}
We evaluate reconstruction performance using the \emph{Levenshtein score}, defined as one minus the normalized edit (Levenshtein) distance.
The score ranges from 0 to 1, with higher values indicating better reconstruction.
The Levenshtein distance is the minimum number of character edit operations (insertions, deletions, and substitutions) required to transform one string into another.
Given predicted and ground-truth strings $s_\mathrm{pred}$ and $s_\mathrm{orig}$,
we evaluate reconstruction accuracy using the score defined as follows:
\begin{equation} \label{eq:levenshtein}
    r_{\mathrm{lev}}(s_\mathrm{pred}, s_\mathrm{orig})
    = 1 - \frac{d(s_\mathrm{pred}, s_\mathrm{orig})}
    {\max\!\left(|s_\mathrm{pred}|, |s_\mathrm{orig}|\right)},
\end{equation}
where $|s_\mathrm{pred}|$ and $|s_\mathrm{orig}|$ denote the string lengths in characters, and $d(s_\mathrm{pred}, s_\mathrm{orig})$ denotes the Levenshtein distance.

\subsection{Comparison with Existing Methods}
\label{sec:exp-limited}
\label{sec:exp-multiplatform}

This comparison evaluates whether the proposed method reconstructs text accurately from fewer keystrokes than existing unsupervised decoders.
We compare the complete proposed pipeline with HMM-based~\cite{zhuang05:keyboard-acoustic} and dictionary-based~\cite{berger2006dictionary} decoders.
The dictionary decoder matches cluster sequences to candidate words through repetition patterns and enforces consistent cluster--character mappings across words.
For both conventional decoders, we apply the same LLM-based correction (Gemini) used in the proposed pipeline to their decoded strings; the HMM-based and dictionary-based curves therefore report scores after LLM correction.

\begin{figure}
    \centering
    \includegraphics[width=0.85\linewidth]{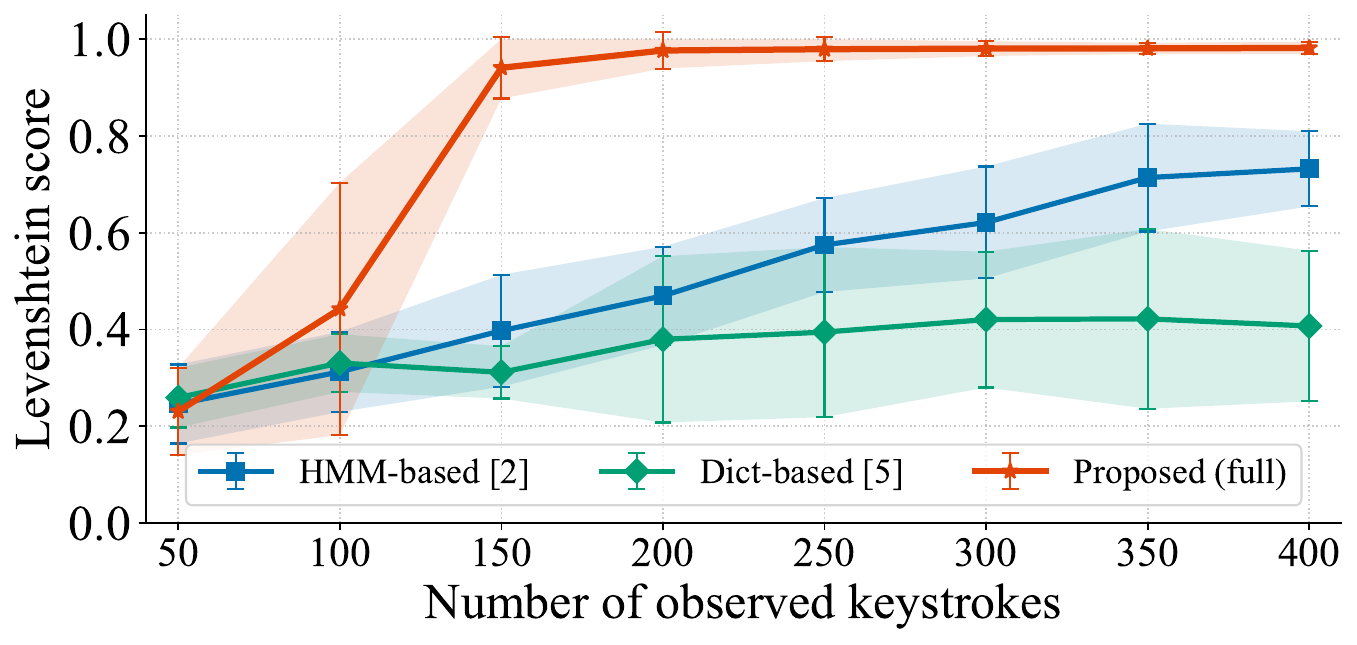}
    \caption{\okada{Comparison with existing decoders.}}
    \label{fig:decoder_comparison_methods}
\end{figure}

All three methods are evaluated under the same recording conditions and acoustic preprocessing configuration.
For each of the 12 device--text recordings in \cref{sec:eval-setup}, we reconstruct text from the first $N$ keystrokes, varying $N$ from 50 to 400 in increments of 50.
All methods use the same 902-dimensional acoustic features (\cref{sec:feature}), UMAP configuration (\cref{sec:umap}), and hierarchical-clustering procedure (\cref{sec:clustering}).
In \cref{fig:decoder_comparison_methods}, each point represents the mean Levenshtein score across the four laptops and three texts.
The error bars and shaded bands indicate one sample standard deviation above and below the mean across these 12 combinations.
They describe variation across devices and texts.

\cref{fig:decoder_comparison_methods} shows the results of the comparison.
The proposed method achieves high reconstruction accuracy with substantially fewer observed keystrokes than either conventional decoder.
At $N=150$, its mean score reaches 0.941, compared with 0.397 for HMM-based decoding and 0.311 for dictionary-based decoding.
At $N=200$, the proposed method improves to 0.977, while the two baselines achieve 0.470 and 0.380, respectively.
Even at $N=400$, the baseline scores reach only 0.732 and 0.407, both below the proposed method's score at $N=150$.
Thus, under the evaluated conditions, the proposed pipeline requires fewer observations for accurate text reconstruction, even when conventional decoders receive LLM-based post-processing.

\subsection{Ablation Study}
\label{sec:eval-ablation}

The ablation study evaluates how iterative feedback and LLM guidance contribute to the proposed method.
We compare BERT decoding alone (BERT only), BERT with iterative feedback but without the LLM (BERT + FB), and the complete pipeline (Proposed (full)).
The BERT-only variant uses the initial BERT prediction without feedback or LLM correction.
BERT + FB treats all BERT predictions as pseudo-labels for label spreading, instead of selecting them through agreement with an LLM-corrected sequence as in \cref{sec:labelspreading}.
The complete pipeline incorporates LLM correction both to guide pseudo-label selection during feedback and to refine the reconstructed text.
All three variants use the same 12 recordings and observation counts as \cref{sec:exp-limited}.
In \cref{fig:decoder_comparison_bert}, points show the mean score, and error bars and shaded bands show one sample standard deviation across the 12 device--text combinations, with shaded bands clipped to the score range.

\begin{figure}
    \centering
    \includegraphics[width=0.85\linewidth]{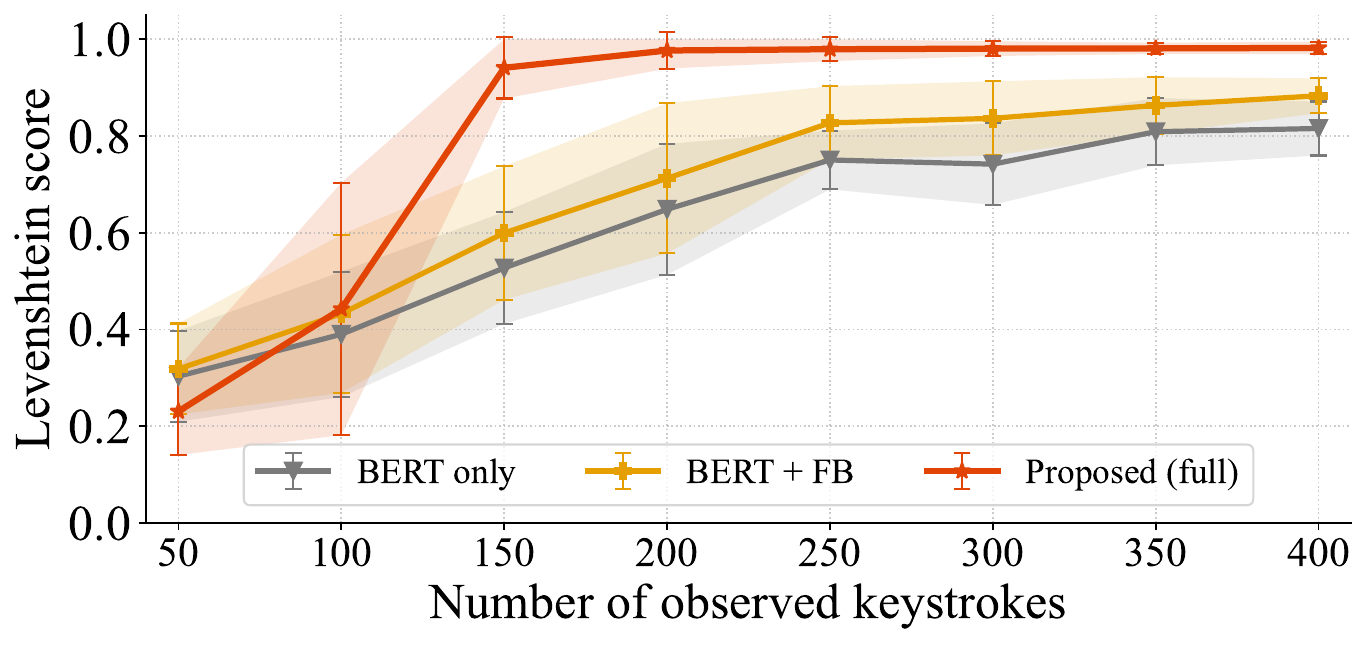}
    \caption{\okada{Ablation of the proposed decoder.}}
    \label{fig:decoder_comparison_bert}
\end{figure}

\cref{fig:decoder_comparison_bert} shows the results of the ablation study.
The BERT-based decoder can reconstruct text without an LLM, and iterative feedback improves its accuracy.
At $N=200$, the BERT-only variant achieves a mean score of 0.648, which increases to 0.713 with feedback.
At $N=400$, these variants reach 0.815 and 0.883, respectively.
These results show that acoustic label spreading and updates to the cluster--character mapping can improve reconstruction using BERT predictions alone.

LLM guidance further improves accuracy and stabilizes reconstruction with fewer observations once sufficient acoustic and linguistic evidence is available.
At $N=150$, the complete pipeline achieves 0.941, compared with 0.527 for the BERT-only variant and 0.599 for BERT + FB.
Its mean score rises to 0.977 at $N=200$ and remains near 0.98 thereafter, while the narrower error bars indicate less variation across device--text combinations.
This benefit does not extend uniformly to the smallest observation counts: at $N=100$, the complete pipeline achieves only 0.442 and exhibits substantial variation.
Together, the results show that an LLM is not required for the decoder to operate, but its correction and feedback guidance substantially improve reconstruction accuracy and stability from approximately 150 observed keystrokes in this evaluation.

\section{Cross-Platform Evaluation under Challenging Acquisition Conditions}
\label{sec:real}

Following the controlled nearby-recording evaluation in \cref{sec:exp-multiplatform}, this section evaluates the proposed attack across the same four laptop platforms and three input texts under three challenging acquisition scenarios: distant recording on the same desk (\cref{sec:exp-distant-desk}), through-the-wall recording (\cref{sec:exp-through-wall}), and online meeting audio (\cref{sec:exp-online-meeting}).
These scenarios comprise five acquisition conditions: desk-surface recording, through-the-wall recording, Google Meet, Microsoft Teams, and Zoom.
The cross-platform evaluation uses 60 recordings across five acquisition conditions.
Across all 60 recordings, the 25{,}800 inter-keystroke intervals had a mean of approximately 368~ms (standard deviation: 151~ms) and a median of approximately 325~ms.
The median corresponds to approximately 185~CPM and falls within the typical 150--300~CPM range considered in \cref{sec:threat-model}.
Under these observed typing conditions, individual keystrokes could be segmented without substantial overlap, satisfying the condition required by our segmentation-based pipeline.

The contact microphone used in \cref{sec:exp-distant-desk,sec:exp-through-wall} captures structure-borne vibrations from the desk or wall rather than airborne sound.
Such vibrations can propagate through rigid structures, enabling keystroke acquisition at larger distances with less sensitivity to ambient acoustic noise.
We use an FL-1000 contact microphone (Sun-Mechatronics)~\cite{sunmechatronics-fl1000}.

\subsection{Distant Recording on Same Desk}
\label{sec:exp-distant-desk}

\begin{figure*}[t]
    \addtocounter{figure}{1}
    \centering
    \includegraphics[width=\linewidth]{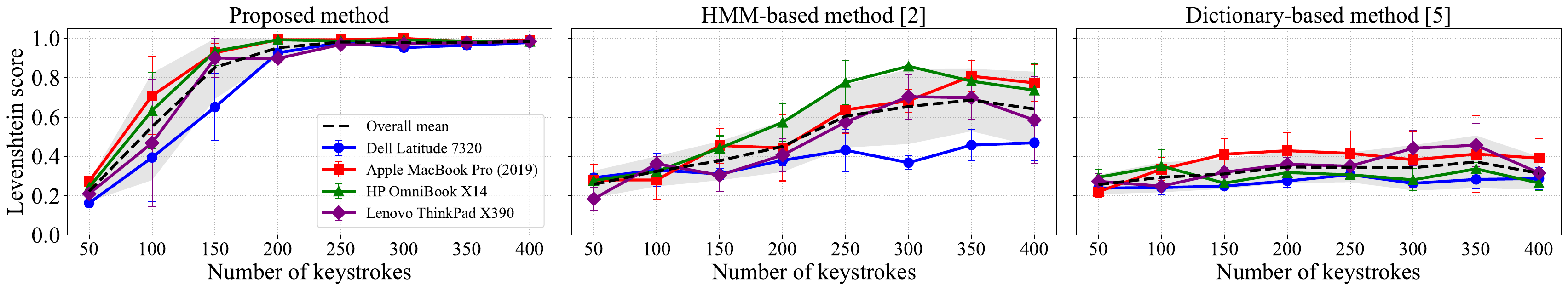}
    \caption{Reconstruction accuracy for distant-on-desk recording.}
    \label{fig:score_concrete}
\end{figure*}

\begin{figure}[t]
    \addtocounter{figure}{-2}
    \centering
    \subfloat[Experimental setup.]{\includegraphics[height=31mm,keepaspectratio]{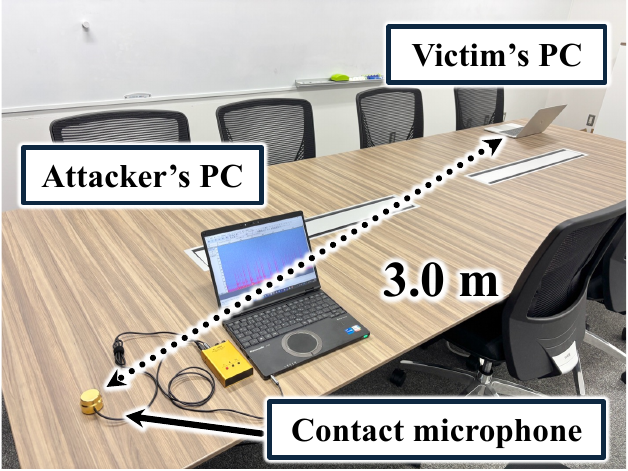}%
    \label{fig:exp2-1_setting_a}}
    \hfil
    \subfloat[Attacker's equipment.]{\includegraphics[height=31mm,keepaspectratio]{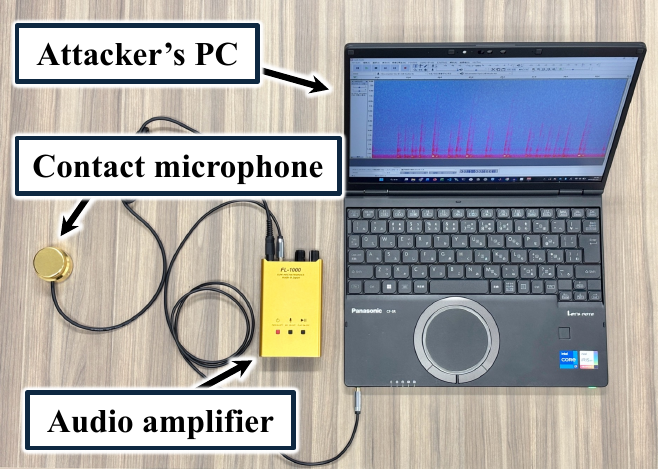}%
    \label{fig:exp2-1_setting_b}}
    \caption{Overview of distant keystroke recording on the same desk.}
    \label{fig:exp2-1_setting}
    \addtocounter{figure}{1}
\end{figure}

We next test whether keystroke information remains exploitable when the attacker records structure-borne vibrations from 3.0\,m away on the same desk.
\cref{fig:exp2-1_setting_a} shows an overview of the experimental setup, and \cref{fig:exp2-1_setting_b} shows the attacker's equipment, which comprises a contact microphone, an audio amplifier, and a laptop used to record the signal.
Although the attacker is depicted using a laptop for recording in \cref{fig:exp2-1_setting_b}, the FL-1000 contact microphone used in our experiments is equipped with an onboard recording function; therefore, a separate recording device is unnecessary.
As in \cref{sec:exp-multiplatform}, we plot the mean score across the three texts.

\Cref{fig:score_concrete} compares the three decoders under distant-on-desk recording.
The proposed method reaches approximately 0.9 or higher across all four laptops by $N=200$, whereas the HMM-based method remains platform-dependent and the dictionary-based method stays substantially less accurate even at $N=400$.
Thus, structure-borne recordings retain sufficient information for accurate reconstruction, but conventional decoders do not exploit it consistently.

\subsection{Through-the-Wall Keystroke Eavesdropping}
\label{sec:exp-through-wall}

\begin{figure*}[t]
    \addtocounter{figure}{1}
    \centering
    \includegraphics[width=\linewidth]{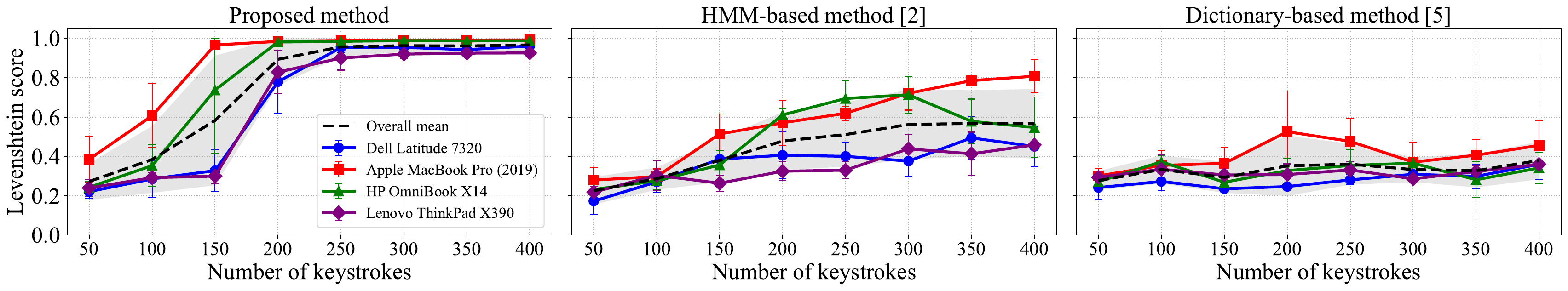}
    \caption{Reconstruction accuracy for through-the-wall eavesdropping.}
    \label{fig:score_wall}
\end{figure*}

\begin{figure}[t]
    \addtocounter{figure}{-2}
    \centering
    \includegraphics[width=0.90\linewidth]{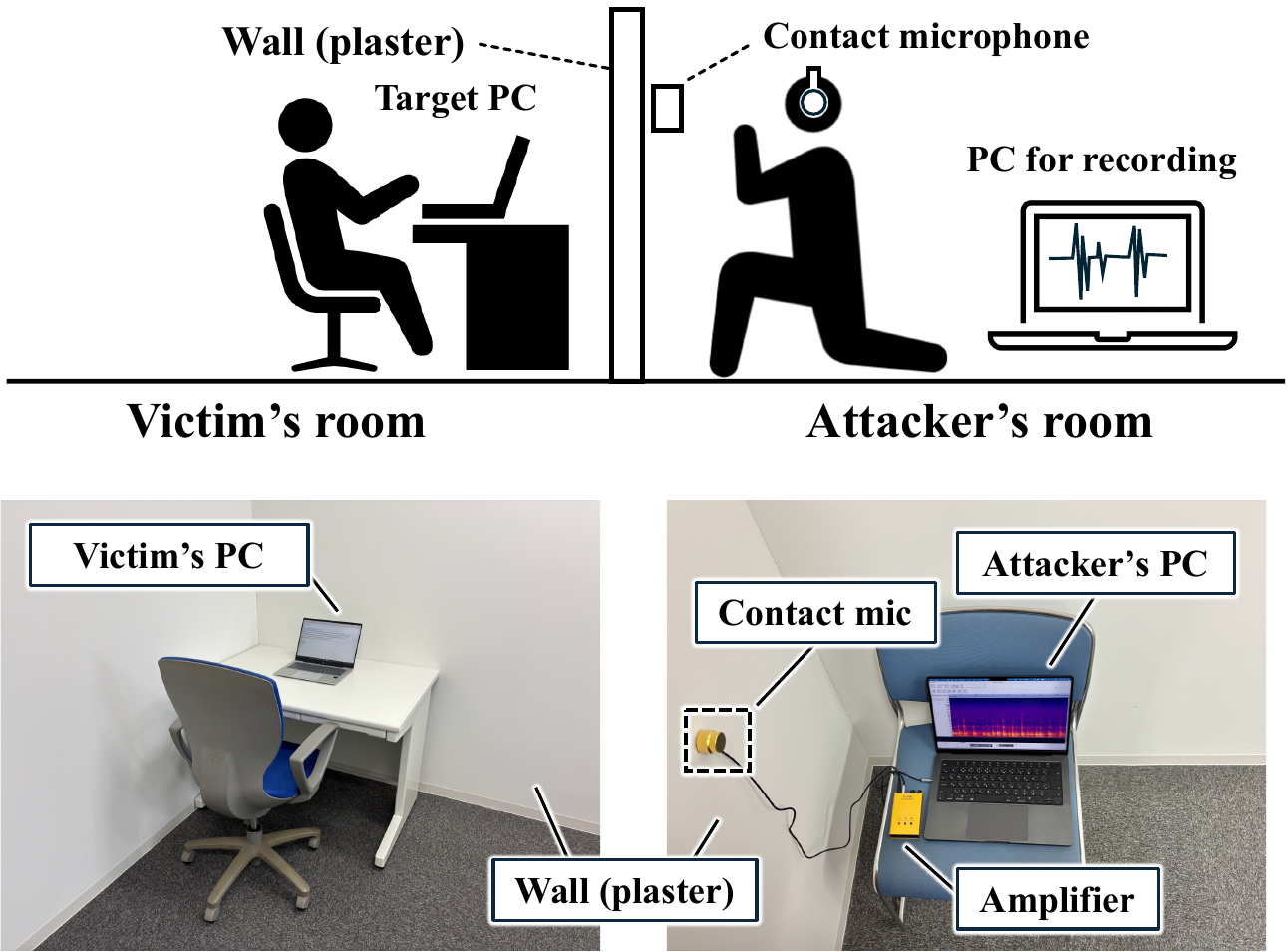}%
    \caption{Overview of through-the-wall keystroke eavesdropping.}
    \label{fig:exp2-2_setting}
    \addtocounter{figure}{1}
\end{figure}

We next test whether structure-borne keystroke information remains exploitable through a wall, without placing the sensor in the victim's room or on the victim's desk.
\cref{fig:exp2-2_setting} shows the arrangement of the attacker and the target across the wall.
For the target-side setup, a laptop is placed on a desk positioned near the wall.
The attacker-side setup consists of a wall-mounted contact microphone, an amplifier, and a laptop for recording.
As in \cref{sec:exp-multiplatform}, we use the same laptop platforms and run reconstruction using only the first $N$ keystrokes, varying $N$ from 50 to 400 in increments of 50.
We plot the mean Levenshtein scores across the three texts. 

\Cref{fig:score_wall} shows that the proposed method reaches approximately 0.9 or higher with $N=150$--$250$, depending on the laptop, despite attenuation and environmental noise introduced by the wall.
In contrast, neither conventional decoder achieves consistently high accuracy across the four laptops by $N=400$.
The decoder advantage therefore persists even when the acoustic observations are obtained through a substantially more challenging propagation path.

\subsection{Attacks via Online Meeting Audio Streams}
\label{sec:exp-online-meeting}

We evaluate whether keystroke sounds transmitted through commodity online-meeting audio remain sufficient for accurate text reconstruction.
We test three widely used conferencing platforms: Google Meet, Microsoft Teams, and Zoom\@.
For each platform, the victim participates in an online call and types while the attacker, joining as another participant, records the received meeting audio.
We use browser-based clients for Google Meet and Microsoft Teams, and a desktop application for Zoom\@.
To isolate the feasibility of the attack under standard audio transmission, we disable noise-suppression and noise-cancellation features in the meeting software, because strong noise suppression can significantly attenuate keystroke sounds.
We emphasize that this condition is not a narrow corner case.
Keystroke-oriented noise suppression is often gated behind specific product tiers or paid subscriptions.
For example, Google Meet distinguishes between device- and cloud-based noise cancellation, with availability depending on the device and Google Workspace configuration, while Studio Sound is restricted to eligible plans and operates together with noise cancellation~\cite{google-meet-noise-cancellation}.
Consequently, keystroke leakage may remain relevant in meeting configurations where effective noise suppression is unavailable or disabled.
Accordingly, our evaluation focuses on meeting configurations in which keystroke information remains present in the transmitted audio.

\begin{figure*}[t]
    \centering
    \subfloat[Google Meet.]{\includegraphics[width=\linewidth]{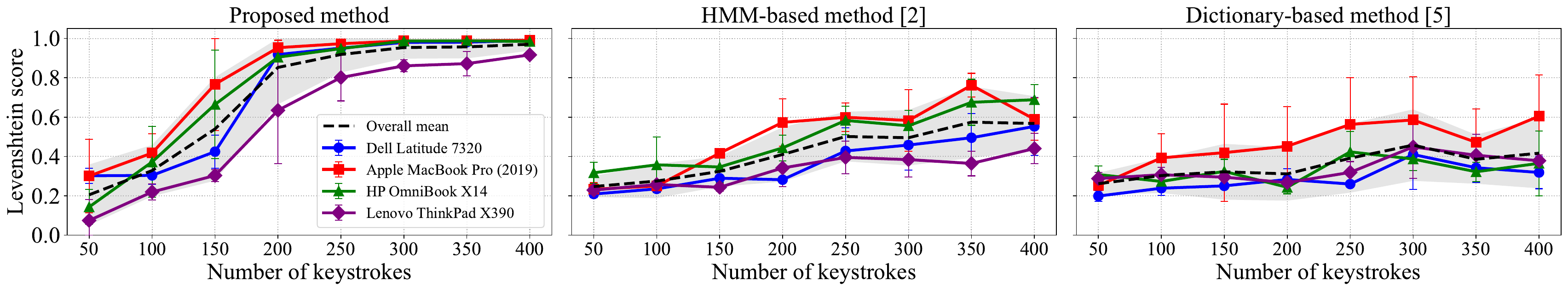}\label{fig:score_meet}}\\
    \subfloat[Microsoft Teams.]{\includegraphics[width=\linewidth]{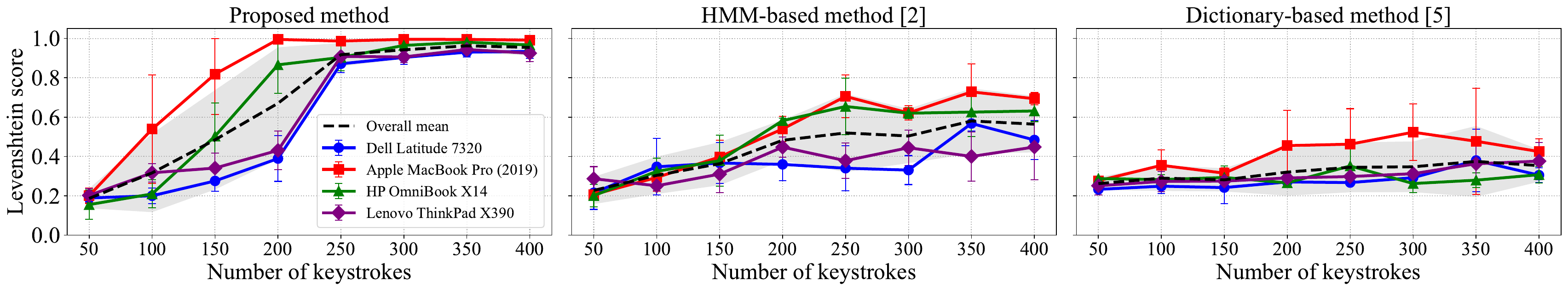}\label{fig:score_teams}}\\
    \subfloat[Zoom.]{\includegraphics[width=\linewidth]{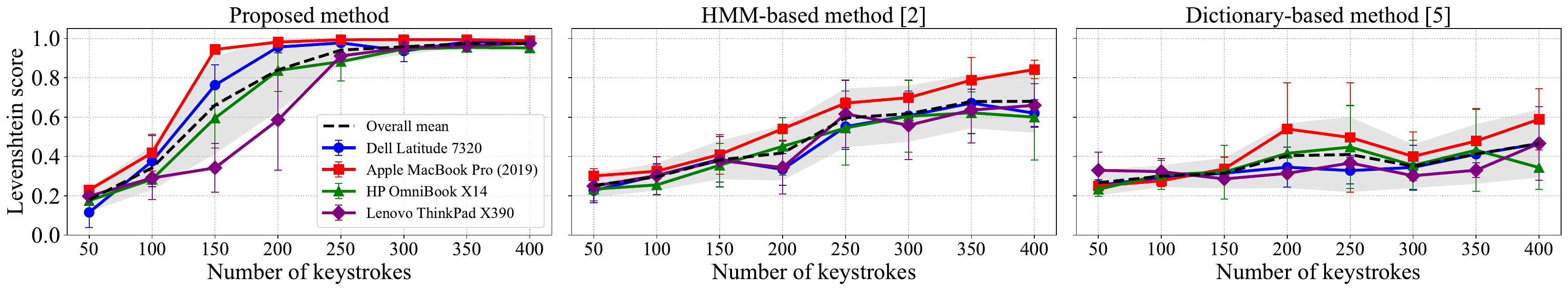}\label{fig:score_zoom}}%
    \caption{Reconstruction accuracy for online-meeting audio streams.}
    \label{fig:score_online_meetings}
\end{figure*}

\Cref{fig:score_meet,fig:score_teams,fig:score_zoom} compare the decoders on Google Meet, Microsoft Teams, and Zoom, respectively.
Across the three services, the proposed method generally reaches approximately 0.9 or higher with $N=200$--$250$, although the Lenovo laptop on Google Meet requires more observations; the HMM- and dictionary-based methods remain substantially lower and more platform-dependent.

Together with the controlled nearby-recording results in \cref{sec:exp-multiplatform} and the distant-on-desk and through-the-wall results in this section, these experiments show that conventional decoders systematically underuse the information retained across practical acquisition channels, whereas the proposed decoder uses that information to reconstruct text reliably from limited observations.

\section{Discussion}
\label{sec:discussion}
In this section, we discuss three further aspects of the proposed attack: robustness to background noise, support for editing and modifier keys, and available mitigations.

\subsection{Noise Robustness}
\label{sec:noise-robustness}
Because practical recordings may contain substantial ambient noise, we evaluate how reconstruction accuracy degrades under controlled noise.
The experiments in \cref{sec:eval,sec:real} used relatively quiet conditions with a sound pressure level (SPL) of approximately 35~dB; here, we conduct an SPL-based evaluation in which physical background noise is played back at a controlled SPL during recording.
In this evaluation, we consider two representative noise types: Gaussian noise and speech-babble noise, the latter consisting of babble-like conversational sound obtained from the MUSAN corpus~\cite{musan} (specifically, \path{musan/noise/free-sound/noise-free-sound-0001.wav}).

Physical background noise is played back through a loudspeaker in the experiment room while keystroke sounds are recorded simultaneously with the three setups of \cref{sec:real}: distant-on-desk recording, through-the-wall recording, and online meeting (Google Meet) audio.
We evaluate three noise levels, 40, 50, and 60~dB SPL, adjusting the loudspeaker volume so that the SPL measured at the target laptop's position matches the intended level; SPL is measured as a 30-s average using a digital sound level meter~\cite{400TST9340:online}.
We use the 13-inch MacBook Pro (2019) and an English passage of 426 characters, both drawn from the experimental configurations used in \cref{sec:eval,sec:real}, as a representative device--text configuration.
We report results for $N=200, 300,$ and $400$ observed keystrokes.

\cref{tab:spl} reports the reconstruction accuracy for each channel, noise type, and SPL.
Scores of 0.950 or higher are shown in bold.
For the distant-on-desk recording, the score stays high under Gaussian noise across all SPLs and under speech-babble noise up to 50~dB SPL; only speech-babble noise at $60$~dB causes a clear drop.
The online-meeting recording shows a similar trend, remaining robust to Gaussian noise at all levels and to speech-babble noise except at the highest SPL with the fewest observations ($N=200$ at $60$~dB).
In contrast, the through-the-wall recording degrades even at the lower noise levels (40--50~dB SPL) and is unstable across observation counts, making it the least robust channel.

\begin{table*}[t]
    \centering
    \caption{\okada{Reconstruction accuracy under controlled background noise at different sound pressure levels.}}
    \label{tab:spl}
    \footnotesize
    \setlength{\tabcolsep}{3pt}
    \begin{tabular}{c|cccccc|cccccc|cccccc}
    \hline
    & \multicolumn{6}{c|}{Distant-on-desk (3\,m)} 
    & \multicolumn{6}{c|}{Through-the-wall}
    & \multicolumn{6}{c}{Online (Meet)} \\
    \cline{2-7}\cline{8-13}\cline{14-19}
    SPL & \multicolumn{2}{c}{$N{=}200$} & \multicolumn{2}{c}{$N{=}300$} & \multicolumn{2}{c|}{$N{=}400$}
    & \multicolumn{2}{c}{$N{=}200$} & \multicolumn{2}{c}{$N{=}300$} & \multicolumn{2}{c|}{$N{=}400$}
    & \multicolumn{2}{c}{$N{=}200$} & \multicolumn{2}{c}{$N{=}300$} & \multicolumn{2}{c}{$N{=}400$} \\
    (dB) & Gauss & Babble & Gauss & Babble & Gauss & Babble
        & Gauss & Babble & Gauss & Babble & Gauss & Babble
        & Gauss & Babble & Gauss & Babble & Gauss & Babble \\
    \hline
    40 & \textbf{0.995} & \textbf{0.990} & \textbf{0.993} & \textbf{0.993} & \textbf{0.975} & \textbf{0.998}
       & 0.605 & 0.258 & \textbf{0.973} & 0.263 & \textbf{0.993} & 0.880 
       & \textbf{0.975} & \textbf{0.975} & \textbf{0.980} & \textbf{0.993} & \textbf{0.990} & \textbf{0.993} \\
    50 & \textbf{0.990} & \textbf{0.985} & \textbf{0.993} & \textbf{0.963} & \textbf{0.990} & \textbf{0.990} 
       & 0.270 & 0.280 & 0.277 & 0.182 & 0.226 & 0.315 
       & \textbf{0.985} & \textbf{0.980} & \textbf{0.990} & \textbf{0.980} & \textbf{0.998} & \textbf{0.990} \\
    60 & \textbf{0.975} & 0.365 & \textbf{0.970} & 0.383 & \textbf{0.973} & 0.260 
       & 0.253 & 0.285 & 0.257 & 0.240 & 0.243 & 0.265 
       & \textbf{0.990} & 0.230 & \textbf{0.983} & \textbf{0.970} & \textbf{0.985} & \textbf{0.998} \\
    \hline
    \end{tabular}
\end{table*}

\subsection{Extending BERT-Based Decoding to Shift and Backspace}
\label{sec:operation-aware}
We next test whether the same self-supervised decoder can extend beyond printable characters to infer Shift and Backspace operations.
The core evaluations in \cref{sec:eval,sec:real} use the basic 29-symbol vocabulary defined in \cref{sec:threat-model} to enable a controlled comparison with the conventional unsupervised baselines, which do not model modifier or editing operations.
We extend this vocabulary with Shift and Backspace using the controlled setup of \cref{sec:eval-setup}.

\paragraph{Operation-aware vocabulary and training}
We extend the 29-symbol vocabulary with three operation tokens, \texttt{[shiftdown]}, \texttt{[shiftup]}, and \texttt{[backspace]}, yielding a total of 32 target tokens.
To generate training sequences for the Shift tokens, we retain capitalization in the OpenWebText corpus, convert each run of uppercase letters to lowercase, and enclose it with a Shift-press token before and a Shift-release token after.
Because ordinary text corpora do not retain erased keystrokes, we generate Backspace training sequences by synthetically injecting typographical errors at a rate of 3\% per character and following them with one to six Backspace tokens.
The injected errors comprise QWERTY-neighbor substitutions, duplicated keystrokes, and random substitutions.
As with the basic model, this training uses only a text corpus and requires no labeled acoustic samples from the target device.

\paragraph{Integration into the reconstruction pipeline}
We extend the cluster--token mapping matrix $\mathbf{M}$ from the basic 29-symbol vocabulary to the 32-token vocabulary.
The BERT decoder and feedback loop then learn the clusters corresponding to Shift press, Shift release, and Backspace jointly with ordinary character clusters.
Before LLM-based correction, the decoded Backspace operations are applied and the Shift-delimited sequences are converted into capitalized text.
Thus, the same reconstruction pipeline infers text content, capitalization, and editing operations without separate heuristic detectors.

\paragraph{Experimental setup and metrics}
We use the same nearby-recording setup as in \cref{sec:eval-setup}, comprising a 13-inch MacBook Pro (2019) and an Apple iPhone 15 placed beside it, and record a modified version of Text~1 containing capitalization and typographical corrections; the full input sequence is provided in Section~S-1 of the supplementary material.
The recording contains 496 keystroke segments: 426 characters in the final text, 23 erroneous characters subsequently deleted, 23 Backspace keystrokes, and 24 Shift events comprising 12 presses and 12 releases.
For each operation token, we report position-wise precision, recall, and F1 score.
We also evaluate the visible text obtained after applying the decoded operations using only the character-level Levenshtein score defined in \cref{eq:levenshtein}.

\paragraph{Results}
\Cref{tab:operation_detection} reports the operation-token detection results after the final feedback iteration.
The extended pipeline detects all 12 Shift presses and all 12 Shift releases without false positives.
It also detects all 23 Backspace keystrokes, with three false positives, yielding a precision of 0.885, a recall of 1.000, and an F1 score of 0.939.
The character-level Levenshtein score improves from 0.611 before feedback to 0.981 after the final feedback iteration; after applying the decoded operations and LLM-based correction, the final visible text achieves a score of 0.972.
These results show that the proposed decoder is not restricted to printable characters: vocabulary expansion and text-only training allow its self-supervised feedback loop to learn clusters for modifier and editing operations without labeled acoustic data or separate key-specific detectors.

\begin{table}[t]
\centering
\caption{Detection accuracy for operation tokens after feedback.}
\label{tab:operation_detection}
\begin{tabular}{l|ccc|ccc}
   \hline
   Operation & TP & FP & FN & Precision & Recall & F1 \\
   \hline
   Shift press   & 12 & 0 & 0 & 1.000 & 1.000 & 1.000 \\
   Shift release & 12 & 0 & 0 & 1.000 & 1.000 & 1.000 \\
   Backspace     & 23 & 3 & 0 & 0.885  & 1.000 & 0.939 \\
   \hline
\end{tabular}
\end{table}

\subsection{Defense and Mitigation}
\label{sec:defence}
The results of this study highlight the need for mitigation strategies that reduce keystroke information leakage across different sensing channels.
Because the proposed attack relies on passive observation of acoustic and vibration signals, effective defenses should address both airborne sound and structure-borne transmission.

\paragraph{Mitigation for online meetings}
Audio processing techniques, such as noise suppression and typing-noise reduction, can significantly reduce the observability of keystroke sounds.
Our evaluation assumes a worst-case scenario in which such processing is disabled or ineffective; when strong suppression is applied and keystroke signals are attenuated, the attack becomes less practical.
This suggests that conferencing platforms and operating systems should enable robust noise suppression by default, provide clearer user feedback when such features are disabled, and support policies that enforce privacy-preserving audio configurations in enterprise environments.
Importantly, these protections should be broadly available, as limiting them to specific product tiers may leave some users exposed to realistic attack risks.

\paragraph{User behavioral mitigation}
Users can reduce exposure by muting microphones or using push-to-talk while typing, increasing microphone–keyboard distance or using external microphones positioned away from the typing surface, and avoiding sensitive typing while audio transmission is active or in exposed public environments.
In addition, as discussed in~\cref{sec:background}, users should avoid typing sensitive information in public and semi-public spaces---such as cafes, libraries, public transportation (e.g., bullet trains and airplanes), and waiting lounges---where keystroke signals may be unintentionally exposed to nearby observers or recording devices.

\paragraph{Physical mitigation}
In addition to airborne sound, our results demonstrate that keystroke information can propagate through structure-borne vibrations.
Mitigation therefore requires reducing vibration transmission in shared physical media.
For example, using soft desk mats, placing vibration-isolating pads beneath the laptop, or selecting materials that damp vibrations can reduce the signal captured by contact microphones.
While such measures may not eliminate leakage entirely, they can degrade the signal-to-noise ratio and make inference less reliable.

Overall, these observations suggest that mitigating keystroke eavesdropping is not solely a software problem but requires coordinated measures across multiple layers of the system.

\section{Related Work}
\label{sec:related}

Keystroke inference has been studied through acoustic, vibration, electromagnetic, wireless, timing, geometric, and visual side channels. 
We focus primarily on acoustic attacks, which are closest to our threat model, and briefly summarize the other modalities.

\paragraph{Acoustic-signal-based approaches}
Acoustic keystroke attacks can be broadly divided into supervised methods that require labeled device-specific recordings and unsupervised methods that infer keys from unlabeled acoustic structure.
Asonov et al.~\cite{asonov2004keyboard} first demonstrated supervised keystroke inference from acoustic emanations. 
Subsequent supervised studies have improved accuracy using refined acoustic features and decoding~\cite{halevi2012closer}, deep models on spectrograms~\cite{harrison2023practical}, and LLM-based correction~\cite{ayati25:llm-typo-correction}, but generally require labeled recordings from the target device or a closely matched condition.

Unsupervised attacks remove this profiling requirement by clustering acoustically similar keystrokes and decoding the resulting sequence. 
Zhuang et al.~\cite{zhuang05:keyboard-acoustic} use an HMM-based language model, while Berger et al.~\cite{berger2006dictionary} and F\"urst et al.~\cite{furst2025practical} exploit dictionary constraints. 
These methods demonstrate inference without labeled acoustic data, but can require many observations or favorable lexical and recording conditions. 
Our work instead targets the low-observation regime and directly addresses uncertainty in the cluster--character mapping.

Recent work uses an LLM to correct an already decoded character string~\cite{ayati25:llm-typo-correction,maurya2026decker}. 
Our framework likewise uses an LLM for residual correction, but its principal advance lies earlier in the pipeline: a character-level BERT model reasons over ambiguity-aware embeddings before hard acoustic decisions, and its predictions iteratively refine the cluster--character mapping.
DECKER~\cite{maurya2026decker} further learns domain-invariant representations from a labeled multi-keyboard corpus and transfers them to unseen keyboards, whereas our setting assumes no labeled acoustic corpus for either training or target adaptation.

\paragraph{Other side channels}
\okada{Other side channels can also reveal keystrokes, but they rely on sensing assumptions that differ from our audio-only threat model.}
Vibration-based attacks infer input from motion or structure-borne signals, including signals from smartphone accelerometers~\cite{owusu2012accessory} and combined motion--acoustic sensing~\cite{murali2018keyboard}.
Electromagnetic attacks exploit unintended emissions from keyboards and cables~\cite{vuagnoux2009compromising}. 
Wireless approaches infer finger motion from Wi-Fi or related physical-layer measurements~\cite{ali2015keystroke,li2016csi,hu2023password}; Yang et al.~\cite{yang22:wireless-training-free} further propose a training-free approach using language constraints.
Other work exploits inter-keystroke timing~\cite{zhang2009peeping,tahiritajar2025acoustic}, acoustically localizes key positions using geometric information~\cite{liu2015snooping,tu2023auditory,zhu2014context,yu2019indirect}, or infers input from visual observations of hands, mobile-device interaction, or eye movements~\cite{shukla2014beware,yue2015blind,chen2018eyetell}. 

Overall, our work is most closely related to unsupervised acoustic attacks, but shifts the focus from the acoustic front end to the decoder: it tests whether modern contextual inference can recover information that HMM- and dictionary-based decoders leave unresolved under limited observations.

\section{Conclusion}\label{sec:conclusion}

We presented a self-supervised acoustic eavesdropping attack that combines acoustic clustering with ambiguity-aware character-level BERT decoding and iterative feedback, without labeled acoustic data.
Under controlled nearby smartphone recording, it achieved mean Levenshtein scores of 0.941 and 0.977 with 150 and 200 observed keystrokes, respectively, across four laptops and three texts, outperforming HMM- and dictionary-based alternatives even with LLM post-processing.
Ablation results show that the BERT decoder operates without an LLM, while iterative feedback and LLM guidance improve performance in the low-observation regime.

The attack remains effective across multiple laptop platforms and increasingly challenging acquisition scenarios, including distant, through-the-wall, and online-meeting recordings. 
An operation-aware extension also infers capitalization and Backspace operations within the same decoding framework, demonstrating that contextual decoding can extend beyond ordinary character keys. 
Together, these results show that keystroke audio acquired under the evaluated conditions can contain sufficient information to reconstruct natural-language text without target-specific acoustic profiling.

The scope of this work suggests two directions for future work.
First, extending the method beyond English natural-language text will require handling strings with weak or no linguistic structure, such as random passwords, as well as expanding the character set to digits and symbols.
Second, adapting the decoding framework to other keystroke-bearing signals, or combining acoustic and non-acoustic evidence, could broaden its applicability beyond the audio-only threat model considered here.

\appendices

\crefname{appendix}{Appendix}{Appendices}
\Crefname{appendix}{Appendix}{Appendices}
\crefalias{section}{appendix}
\section{Training and Inference Details for the Character-Level BERT}
\label{ap:bert}

We train the character-level BERT using a masked language modeling (MLM) objective on a corpus derived from OpenWebText~\cite{Gokaslan2019OpenWeb}, preprocessed to retain only the 29 target symbols and split into 95\% training and 5\% validation data. 
Each input is tokenized into character IDs, and 15\% of the tokens are randomly replaced with \texttt{[MASK]}. 
The model is trained for 10 epochs using cross-entropy loss, an effective batch size of 8{,}192, and a learning rate of $3\times10^{-4}$. 
Optimization used AdamW ($\beta_1=0.9$, $\beta_2=0.999$, $\epsilon=10^{-8}$) with zero weight decay and no warmup; the learning rate decayed linearly to zero. 
Gradients were clipped to a maximum norm of 1.0, and training was performed in FP32.
\Cref{tab:BERT-Param} summarizes the model architecture. 
Training was performed on a system with dual AMD EPYC 9355 CPUs, eight NVIDIA RTX 6000 Ada GPUs, and 256~GB of memory.

\begin{table}[t]
    \centering
    \caption{Character-level BERT hyperparameters.}%
    \label{tab:BERT-Param}
    \begin{tabular}{l|c||l|c}
    \hline
    Hyperparameter & Value & Hyperparameter & Value\\
    \hline
    Vocabulary size & 29 & \# of Transformer layers & 8 \\
    Hidden size & 384 & \# of Attention heads & 8 \\
    Intermediate size & 1536 & Max sequence length & 256 \\
    \hline
    \end{tabular}
\end{table}

\paragraph{\okada{Inference for long sequences}}
For sequences longer than 256 tokens, we apply BERT to overlapping 256-token windows shifted across the sequence and merge the predictions in overlapping regions.

 
\bibliographystyle{IEEEtran}
\bibliography{references.bib}

\vspace{-33pt}
\begin{IEEEbiography}[{\includegraphics[width=1in,height=1.25in,clip,keepaspectratio]{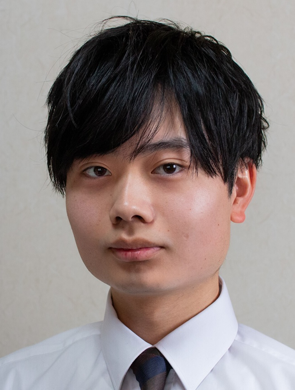}}]{Atsunori Okada}
received the B.E. degree in the Course of Innovative Electrical and Electronic Engineering, National Institute of Technology (KOSEN), Oyama College, Japan, in 2025. He is currently pursuing the master's degree in the Graduate School of Engineering, Tohoku University, Japan. His research interests include hardware security.
\end{IEEEbiography}

\vspace{-33pt}
\begin{IEEEbiography}[{\includegraphics[width=1in,height=1.25in,clip,keepaspectratio]{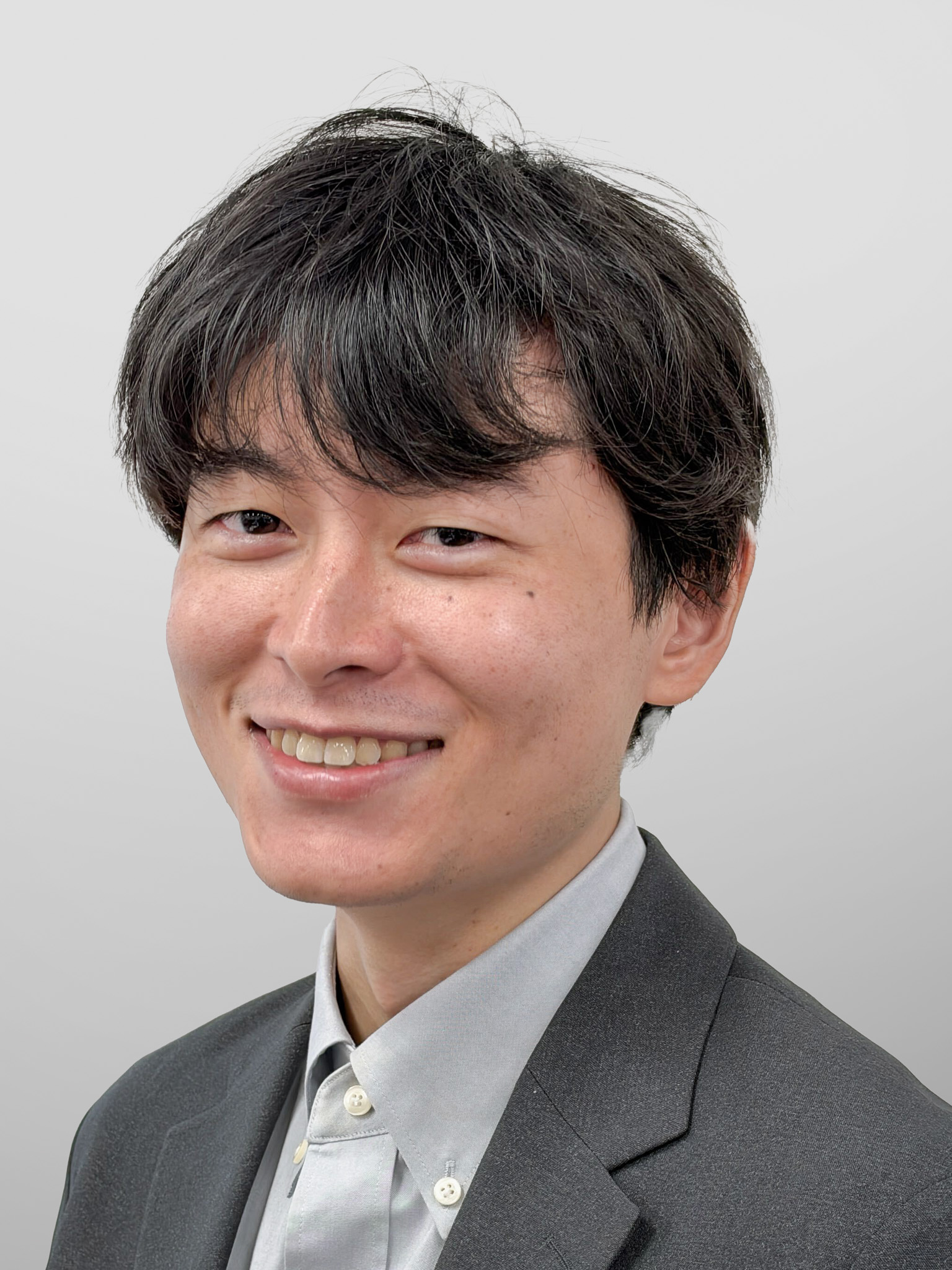}}]{Akira Ito}
received the B.E. degree in information engineering, the M.S. degree in information sciences and the Ph.D. degree in engineering from Tohoku University, Japan, in 2017, 2019, and 2022, respectively. From 2022 to 2025, he was a researcher at NTT Inc. He is currently an Assistant Professor at the Research Institute of Electrical Communication, Tohoku University. His research interests include hardware security and machine learning. He received the Kenneth C. Smith Early Career Award in Microelectronics at ISMVL 2022.
\end{IEEEbiography}

\vspace{-33pt}
\begin{IEEEbiography}[{\includegraphics[width=1in,height=1.25in,clip,keepaspectratio]{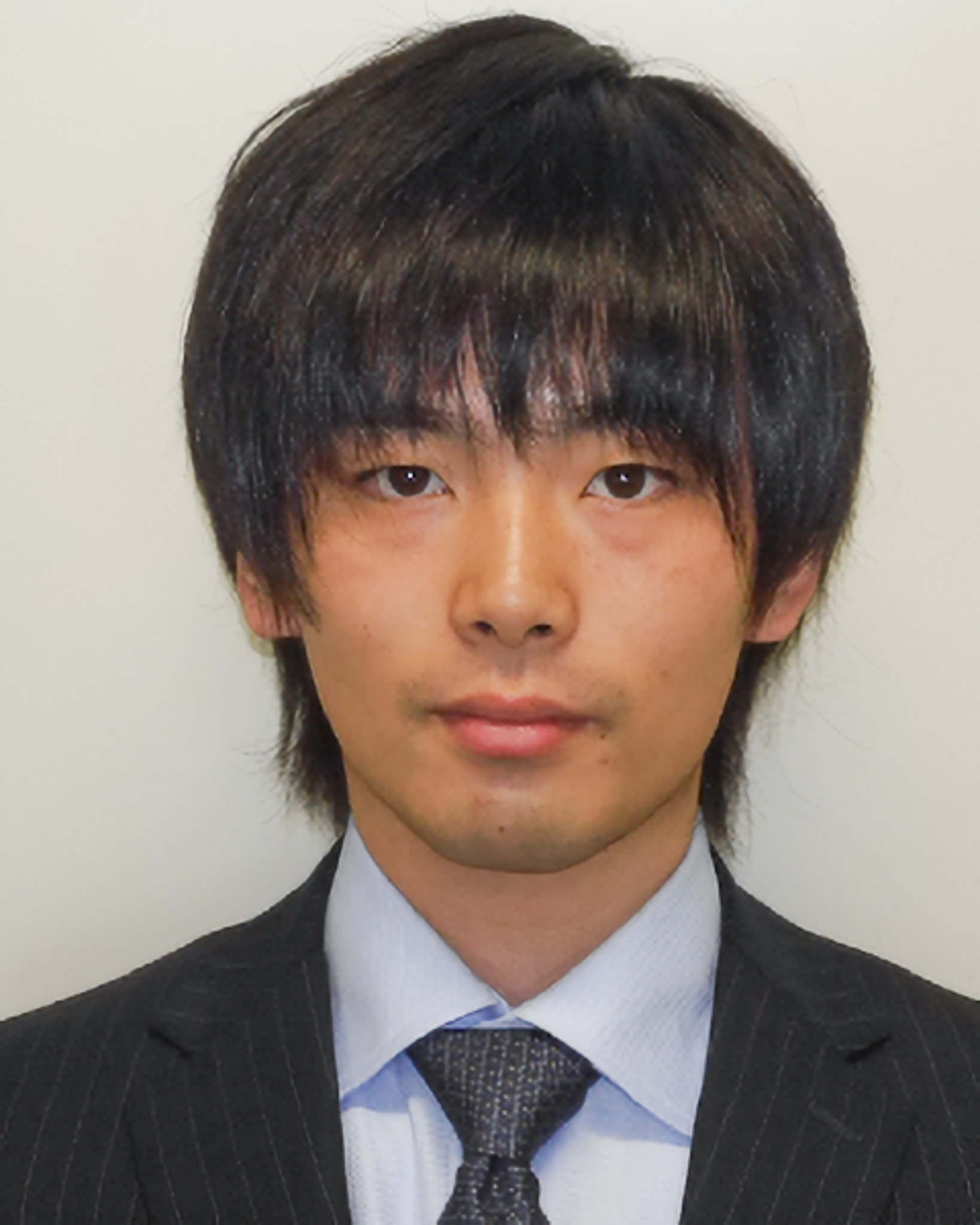}}]{Rei Ueno}
 received the B.E. degree in Information Engineering, and the M.S. and  Ph.D. degrees  in Information Sciences from Tohoku University, Japan, in 2013, 2015, and 2018, respectively.
 He was an assistant professor at the Research Institute of Electrical Communication, Tohoku University, during 2018--2024 and also joined JST as a researcher for a PRESTO project during 2018--2022. He has been an associate professor at the Graduate School of Informatics, Kyoto University, Japan, since 2024. His research interests include theoretical aspects of cryptographic implementations, side-channel attacks, hardware security, secure computer architecture, and information-theoretic cryptography. Dr. Ueno received the Kenneth C. Smith Early Career Award in Microelectronics at ISMVL 2017.
\end{IEEEbiography}

\vspace{-33pt}
\begin{IEEEbiography}[{\includegraphics[width=1in,height=1.25in,clip,keepaspectratio]{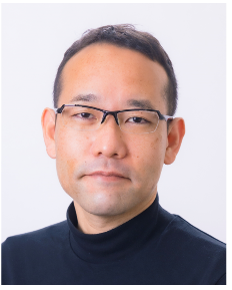}}]{Yuichi Hayashi}
 is a Professor at Nara Institute of Science and Technology. His research interests include electromagnetic compatibility and hardware security. He is the chair of the EM information leakage subcommittee in IEEE EMC Society Technical Committee 5. He serves as a member of the IEEE EMC Society Board of Governors. He has been recognized through many awards and honors, including the IEEE International Symposium on Electromagnetic Compatibility Best Symposium Paper Award (2013), IEEE Electromagnetic Compatibility Society Technical Achievement Award (2021), and the Richard B. Schulz Best Transactions on EMC Paper Award (2024). 
\end{IEEEbiography}

\vspace{-33pt}
\begin{IEEEbiography}[{\includegraphics[width=1in,height=1.25in,clip,keepaspectratio]{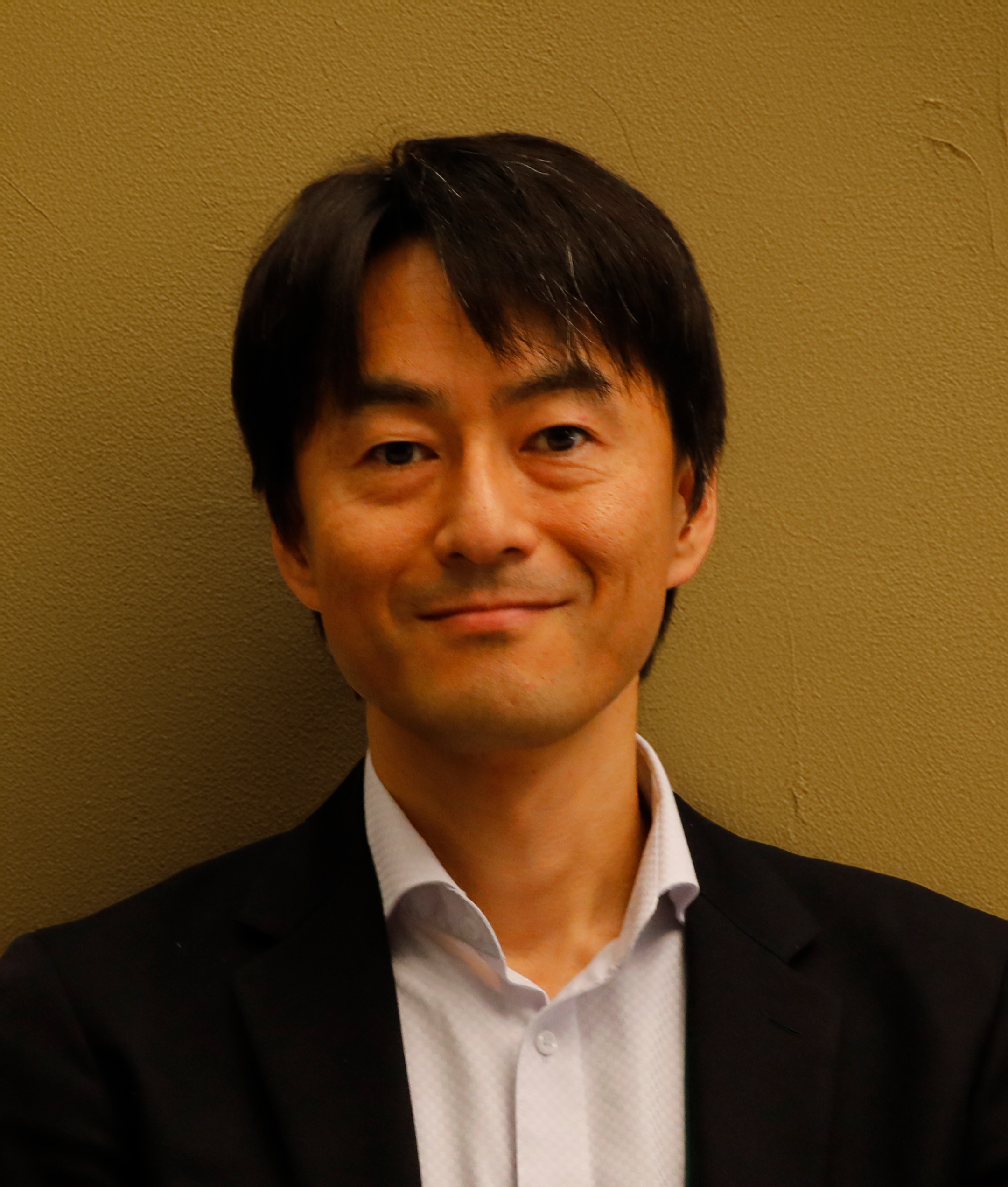}}]{Naofumi Homma}
received the Ph.D. in information sciences from Tohoku University, Japan, in 2001.  Since 2016, he has been a Professor in the Research Institute of Electrical Communication, Tohoku University.  In 2009-2010 and 2016-2017, he was a visiting professor at Telecom ParisTech in France. His research interests include cryptographic computing, system security, electronic design automation, and next-generation security by design. He received more than 30 awards, including the Best Paper Award at the IEEE EMC in 2013, the Best Paper Award at the IACR CHES in 2014, and the Japan Society for the Promotion of Science (JSPS) Prize in 2018. He served as a chairman/member of several academic/industrial committees, including the chair of the IEEE Computer Society Technical Committee on Multiple-Valued Logic, a chair of the JEITA Device and Hardware Security Technical Committee, and a member of the Advisory Board for Cryptography Technology established by the Japanese government.
\end{IEEEbiography}

\vfill

\end{document}